\documentclass[final,5p,times,twocolumn,authoryear]{elsarticle}
\usepackage{amssymb}
\usepackage{amsmath}
\usepackage{newtxtext,newtxmath}
\usepackage[T1]{fontenc}
\usepackage{graphicx}	
\usepackage{ae,aecompl}
\usepackage{ulem}
\usepackage{booktabs}
\usepackage{hyperref}
\usepackage{natbib}
\usepackage{color}
\usepackage{caption}
\usepackage{subcaption}
\usepackage{enumitem}
\usepackage{parskip}
\DeclareRobustCommand{\VAN}[3]{#2}
\let\VANthebibliography\thebibliography
\def\thebibliography{\DeclareRobustCommand{\VAN}[3]{##3}\VANthebibliography}

\newcommand{\orcid}[1]{%
\href{https://orcid.org/#1}{\includegraphics[width=8pt]{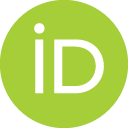}}%
}

\begin{document}

\begin{frontmatter}



\title{Universal Relations and Correlation Analysis of Proto-Neutron Star Properties in Energy-Momentum Squared Gravity}
\author{Sayantan Ghosh\orcid{0000-0001-8276-1935}}
\ead{sayantanghosh1999@gmail.com}

\affiliation{organization={Department of Physics and Astronomy, National Institute of Technology},
            city={Rourkela},
            postcode={769008}, 
            state={Odisha},
            country={India}}
         
\begin{abstract}
Proto-neutron stars (PNSs) are the hot, lepton-rich remnants of the core collapse supernovae, which go through a cooling phase and become cold, stable Neutron stars (NSs). Since PNSs are also superdense objects with strong gravitational fields, we can use them to probe general relativity (GR) in the high-curvature regime, similar to NSs. In this study, we analyze the macroscopic properties like mass, radius, compactness, tidal deformability, $f$-mode oscillations and gravitational binding energy of PNSs using four different relativistic mean-field (RMF) equations of state (EOSs) with fixed entropy per baryon ($S$ =1, 2) and varying the lepton fractions ($Y_l$). The variation of $S$ and $Y_l$ has a noticeable effect on these properties. Extending our study beyond GR, we explore these effects within the framework of Energy-Momentum Squared Gravity (EMSG). This modified gravity theory adds the squared energy-momentum terms to the field equations with a free parameter $\alpha$. In the weak-field regimes, EMSG remains indistinguishable from GR, but in the strong-field regimes, such as PNSs or NSs, it shows measurable deviations. Varying the free parameter $\alpha$, we observe significant changes in the macroscopic properties of the PNSs. After that, we focus on the universal relations of the macroscopic properties and the correlations of the universal relations. We find that, despite significant changes in the macroscopic properties induced by the variations of $S$, $Y_l$ and $\alpha$, the correlations remain strong and nearly unaffected.
\end{abstract}

\end{frontmatter}




\section{Introduction}
\label{intro}

Neutron stars (NSs) harbour an extensive dense matter, roughly several times nuclear density ($n_0\sim 10^{14}gm/cc$), throughout their outer crust to their inner core. That's why it has become a challenge to formulate a nuclear equation of state (EOS) under such diverse conditions. Moreover, in some dynamic scenarios like binary neutron star (BNS) mergers \citep{PhysRevD.83.124008,PhysRevD.86.064032,Kaplan_2014,PhysRevC.96.045806,Lalit2019,Beznogov_2020,PhysRevD.102.043006,PhysRevD.103.083012,PhysRevD.104.063016,Koliogiannis_2021}, core-collapse supernovae \citep{BA2013, CC2014, HTJ2016, M20}, and proto-neutron stars (PNSs) \citep{HS2016, LV2004, Pons_1999, PRAKASH19971, Kumar2020}, the EOS becomes highly temperature dependent. PNSs are extremely hot and lepton-rich, which evolve after the core collapse supernova. So by emitting neutrinos (which were trapped initially \citep{PhysRevD.74.123001, Tripathy2012}) and going through the cooling-down process, PNSs become cold NSs. Throughout the cooling process, the finite-temperature EOS becomes relevant \citep{PhysRevC.106.025801, Raithel_2019, PhysRevD.100.066027, PhysRevC.24.1191, SHEN1998435, GHOSH_FTT} to describe the changes in its structure and composition over time. This phase is crucial in shaping the star's subsequent evolution and the surrounding environment's dynamics. A PNS may transform into a NS or further contract into a black hole (BH), depending on the temperature and other variables. The temperature plays a crucial role in determining the fate of the PNS. If the temperature is high enough, the PNS can retain sufficient thermal energy to prevent further gravitational collapse, leading to the formation of a stable NS. However, if the temperature is too low, the PNS may continue to collapse and eventually produce a BH. The finite-temperature EOS also becomes crucial for NS merger events (as we mentioned before), such as GW170817 \cite{GW170817}. The remnant formed in such events could either be a hypermassive NS or may go through prompt collapse and become a black hole (BH), with a finite-temperature EOS being crucial for an accurate description under conditions of extreme heat and pressure. 
\\
PNSs with their extreme gravitational fields also provide us with an ideal laboratory to study general relativity (GR) in the high curvature regime and to explore possible deviations via modified gravity theories. It has been demonstrated that energy-momentum squared gravity (EMSG) and GR cannot be distinguished using local tests, such as Solar System experiments, because both theories produce identical gravitational potentials in the weak-field regime, leading to the same parametrized post-Newtonian (PPN) parameters \cite{AKARSU2023101305}. Consequently, the motion of test particles, governed by geodesic equations, remains unchanged in both frameworks, making any EMSG-induced deviations undetectable in classical gravity tests. This necessitates alternative approaches—such as NS observations—to probe EMSG effects in strong-field regimes. NSs or PNSs serve as natural laboratories for testing deviations from GR because their extreme densities amplify the effects of modified gravity theories like EMSG. While Solar System tests rely on weak-field approximations, NSs are governed by strong-field gravity, where EMSG-induced modifications become significant \cite{AKARSU2023101305, EMSG_OAkarsu, EMSG_NAlam}.
\\
Following \cite{AkarsuNew}, we analyze EMSG theory compared to GR to examine PNS properties like mass, radius, compactness, non-radial oscillation, binding energy and tidal deformation \cite{Rahmansyah}. While previous EMSG research has focused on its theoretical aspects \cite{Akarsu, EMSG_OAkarsu}, macroscopic NS properties \cite{EMSG_NAlam}, charged quark stars \cite{PRETEL2023169440}, and universal relations (URs) for NSs in EMSG \cite{Ghosh2025}, the URs and their correlations for PNSs in EMSG remain unexplored. This study investigates various URs on PNS tidal deformability, compactness, gravitational binding energy and $f$-mode frequency by varying the $\alpha$ parameter. 
\\
For the current study, we have used 4 different diverse EOSs based on the RMF formalism. The EOS models are NITR \cite{Routaray_2023}, IOPB-I \cite{kumar_2018},  MODEL I \cite{PARA_MODEL1}, and IUFSU \cite{PARA_IUFSU}. All the EOSs used are consistent with current observational constraints of NS maximum mass till date to $\approx$ 2 M$_\odot$. The coupling constants and empirical values of properties associated with nuclear matter at saturation for the parameter sets are given in Table \ref{params}. In \cite{GHOSH_FTT}, we presented the ranges for $E_{\rm sym}$ and $L$ as estimated by various studies, which analyze nuclear data from terrestrial experiments, astrophysical observations, and theoretical calculations \cite{Lattimer_2013, Hagen2015, RocaMaza2015, Oertel2017}. All 4 parameter sets used in this work are consistent with these data. Using 4 EOS models, we generated data by altering the \(\alpha\) parameters. We established URs of $f$- mode at $M_{1.4M\odot}$ ($f_fM_{1.4}$) and binding energy per mass ($B/M$) with compactness ($C$) and dimensionless tidal deformability ($\Lambda$). Finally, we studied the correlation of the above-mentioned URs for PNS. This is the first study to investigate the macroscopic properties of PNSs and the correlation between the URs of PNSs in the context of EMSG.
\\
In this work, we studied the macroscopic properties and the universal relations of PNS during its evolution phase, which will eventually become a NS. We have considered the unified EOS models for S=0 (i.e. for NS), but for PNS (S=1 and S=2) (Unit of S: $k_B$, which is the Boltzmann constant), along with the core part from these models, we have considered HS(DD2) \cite{HEMPEL, DD2} as finite temperature crust. We have extracted the finite temperature crust data from \href{https://compose.obspm.fr}{COMPOSE}.
\\
This paper is organised as follows. In Sec. \ref{sec:TF}, we describe the theoretical framework, which is further detailed in the following subsection. In \ref{TDES} we discuss temperature-dependent EOSs, in \ref{sec:EMSG} we discuss EMSG and in SubSec. \ref{sec:hydrostatics} Hydrostatic equilibrium in EMSG has been presented. We present our results and discussion part in section (\ref{sec:R&D}), which includes the effect of temperature and entropy on NS properties. In section \ref{sec:UR} we study the universal relations. We present the correlation study in section \ref{sec:CorrUR}. And finally, in Section \ref{sec:summary}, we conclude our work.
\\
Throughout this paper, we adopt mostly positive signatures (-, +, +, +) and utilize a geometrized unit system ($G=c=\hbar=1$).
\section{Theoretical Framework}
\label{sec:TF}
In this section, we have presented our detailed theoretical framework related to this study. We first generated the finite temperature dependent EOSs by keeping the entropy per baryon ($S$) constant and varying lepton fraction ($Y_l$). Then we have considered the EMSG theory and obtained modified equations in hydrostatic equilibrium. We then first verify the causality condition by calculating sound speed squared ($c_s^2$). After that, we solve the modified non-radial oscillation equations in Cowling approximation, followed by the calculation of the gravitational binding energy. The following subsections contain the details of the theoretical background.
\subsection{Temperature Dependent Equation of State }
\label{TDES}
To generate the temperature-dependent EOSs, we have incorporated and adapted the formalism from earlier works \citep{Kochankovski_2022}. Previously, we have also done one work based on this formalism \cite{GHOSH_FTT}, where we consider matter composed of baryons and leptons at temperatures $T$, baryon densities $n_{B}$, and lepton fractions $Y_{l}$. The exchange of different mesons serves as a model for the interaction of baryons in the relativistic field theory \cite{WALECKA1974491, BOGUTA1977413, Serot_1997, PhysRevC.90.044305}.
The densities at finite temperature, in scalar $n_B^s$ and vector $n_B$ form (for the baryon
Dirac field $\Psi_B$), are provided by
\begin{equation}
\begin{aligned}
\label{eq:bary_dens}
n_B= & \langle\bar{\Psi}_B \gamma^0 \Psi_B\rangle \\
= & \frac{\gamma_B}{2 \pi^2}\int_0^{\infty}\!\! dk \, k^2 \, (f_{B}(k,T) - f_{\bar{B}}(k,T))  , \hspace{8cm} 
\end{aligned} 
\end{equation}
\begin{equation}
\begin{aligned}
n_B^s= & \langle\bar{\Psi}_B\Psi_B\rangle  \\
= & \frac{\gamma_B}{2 \pi^2}\int_0^{\infty} \!\! dk \, k^2 \, \frac{m^*_B}{\sqrt{k^2+m_B^{*2}}}\left(f_{B}(k,T)+f_{\bar{B}}(k,T)\right), \hspace{8cm}
\end{aligned} 
\end{equation}
\begin{table*}
  \centering
  \setlength{\tabcolsep}{8 pt}
  \caption{Parameters of the model NITR, IOPB-I, MODEL I and IUFSU. The mass of the nucleon is equal to $m_N = 939$ MeV. Also listed in the lower panel are a few nuclear matter properties. We present the saturation density ($n_0$), energy per particle ($E/A$), symmetry energy ($E_{\rm sym}$), slope of the symmetry energy ($L$) at $n_0$ and compressibility ($K$) in symmetric nuclear matter. The empirical/experimental values are given in the last column with their Refs. [a]\cite{Bethe_1971}, [b]\cite{Colo_2014}, [c]\cite{Danielewicz_2014}.}
  \scalebox{1}{\begin{tabular}{ccccccc}
     \hline \hline\noalign{\smallskip}
     & NITR & IOPB-I& MODEL I &IUFSU & Emp./Expt. & Unit \\
      \hline\noalign{\smallskip}
    $m_{\sigma}$ &492.988  &500.512& 496.0067 & 491.5 & &MeV \\
    $m_{\omega}$ &782.5  &782.5& 782.5 &782.5 & &MeV \\
    $m_{\rho}$ &763.0   &763.0& 763.0 & 763.0 & &MeV \\
    $g_{\sigma N}^2$  &97.397   &108.526& 112.8814 &99.4266 & &- \\
    $g_{\omega N}^2$ &160.833   &178.342& 192.0580 &169.8349 &  &- \\
    $g_{\rho N}^2$ &201.413   &123.688& 145.8539 & 184.6877 & &- \\
    $\kappa$ &4.581   &3.616& 3.395 & 3.3808 & &MeV \\
    $\lambda$ &-0.016   &-0.0075 & -0.00239 & 0.000296 & &- \\
    $\zeta$ &0.006   &0.017& 0.0271  & 0.03 & &- \\
    $\Lambda_{\omega}$ &0.045   &0.014& 0.032061 & 0.046 & &- \\
    \hline \smallskip
    $n_0$ & 0.155 & 0.149& 0.150  & 0.155  & $0.148-0.185^{[a]}$ &$fm^{-3}$ \\
    $E/A$ & -16.34 & -16.10& -16.036  & -16.40  & $(-15) - (-17)^{[a]}$ &MeV\\
    $K$ & 225.11 & 222.65& 210.12  & 231.20  & $220-260^{[b]}$ & MeV\\
    $E_{\rm sym}(n_0)$& 31.69 & 33.30& 32.03  & 31.30  & $30.20-33.70^{[c]}$ & MeV\\
    $L$ & 43.86 & 63.58& 57.62 & 47.20  & $35.00-70.00^{[c]}$ & MeV\\
    \noalign{\smallskip}\hline
    \hline
    \end{tabular}}
  \label{params}
\end{table*}
with $\gamma_B=2$ representing the degeneracy of the baryon spin degree of freedom~\citep{kumar_2018,PhysRevC.100.054314,PhysRevC.61.054904} and
\begin{equation}
f_{B/\bar{B}}(k,T) =\left[1+\text{exp}\left(\frac{\sqrt{k^2+m_B^{*2}} \mp \mu_B^{*}}{T}\right)\right]^{-1},
\label{eq:distribution}
\end{equation}
being the Fermi-Dirac distribution for the baryon ($b$) and antibaryon ($\bar b$) with effective mass $m_B^*$ and the accompanying effective chemical potential provided by
\begin{equation}
\mu_B^{*} = \mu_B - g_{b\omega}\bar \omega  - g_{b\rho} I_{3b} \bar \rho- g_{b\phi}\bar \phi.
\label{eq:mueff}
\end{equation}
NS cores are composed of globally charge-neutral matter in $\beta$-equilibrium \cite{SHEN1998435, Stone}.
\\
Therefore, by considering the conditions in the core of the NS related to the chemical potentials and the number densities of the various particles, the appearance of different particles \cite{sedrakian_equation_2021}, and conservations of baryons and leptons, we calculate the overall baryon density \cite{2017IJMPD..2650077Z, Pons_1999, PRAKASH19971,2018arXiv180101350C, PhysRevLett.66.2701}. The appearance of different particles for different baryon densities for these four EOSs has been discussed in \cite{GHOSH_FTT}.
\\
The entropy density is calculated by \cite{Prakash_1997,10.1143/PTP.100.1013}
\begin{align}
s= \sum_i & \frac{\gamma}{2\pi^2} \int_0^{\infty} dk\,k^2\,[-f_{B}(k,T)\ln f_{B}(k,T) \nonumber  \\ 
- & \left(1-f_{B}(k,T)\right)\ln \left(1-f_{B}(k,T)\right) \nonumber \\ 
- & f_{\bar{B}}(k,T)\ln f_{\bar{B}}(k,T) \nonumber \\
- & \left(1-f_{\bar{B}}(k,T)\right) \ln \left(1-f_{\bar{B}}(k,T)\right)].
\label{Eq:entropy}
\end{align}
We have defined the entropy per baryon (S) as S=$s$/n, where n is the baryon number density.
\\
The energy density and pressure of the NS matter can be calculated from the chemical potential and species density at a given $n$ \cite{PhysRevC.89.044001, PhysRevD.104.063016}.
From the stress-energy momentum tensor ($T^{\mu \nu}$), the other thermodynamic variables can be easily calculated, like the energy density $\mathcal{E} = \langle T^{00}\rangle$, and the pressure $P = \frac{1}{3}\langle T^{jj}\rangle$ \cite{GHOSH_FTT, Fetter, Kochankovski_2022}.

\subsection{Energy-Momentum Squared Gravity (EMSG)}
\label{sec:EMSG}
In modified gravity, one extra term as a function of $\mathcal{L}_{\mathrm{m}}$ \cite{Harko2010}, $g_{\mu\nu}T^{\mu\nu}$ \cite{Harko2011}, or $T_{\mu\nu}T^{\mu\nu}$ \cite{Katırcı2014,PhysRevD.94.044002,PhysRevD.97.024011,PhysRevD.96.123517}, represented as $f (\mathcal{L}_{\mathrm{m}}, g_{\mu\nu}T^{\mu\nu}, T_{\mu\nu}T^{\mu\nu})$ is being added in the gravitational action and we get the modified action as \cite{AkarsuNew}: 
\begin{align}
S=\int \left[\frac{1}{2\kappa}\mathcal{R}+f (\mathcal{L}_{\mathrm{m}}, g_{\mu\nu}T^{\mu\nu}, T_{\mu\nu}T^{\mu\nu})+ \mathcal{L}_{\mathrm{m}}\right]\sqrt{-g}\,\mathrm{d}^4x,
\label{action}
\end{align}
All the terms have their usual meaning \cite{Ghosh2025, ghosh2025curv}. For EMSG, the function $f$ is defined as a quadratic function of  $T_{\mu \nu }$, as $f(\mathcal{L}_{\mathrm{m}}, g_{\mu\nu}T^{\mu\nu}, T_{\mu\nu}T^{\mu\nu})=\alpha T_{\mu\nu}T^{\mu\nu}$ \cite{Ghosh2025, ghosh2025curv, Katırcı2014, PhysRevD.94.044002, PhysRevD.97.024011, PhysRevD.96.123517, EMSG_OAkarsu, EMSG_NAlam}. The strength of the EMSG correction is governed by the term $T_{\mu \nu }T^{\mu \nu }$, which is scaled by a real constant parameter $\alpha$.
The energy-momentum tensor ($T_{\mu\nu}$) in terms of $\mathcal{L}_{\mathrm{m}}$, is defined as \cite{EMSG_OAkarsu,EMSG_NAlam}
\begin{align}  \label{tmunudef}
T_{\mu\nu}=-\frac{2}{\sqrt{-g}}\frac{\delta(\sqrt{-g}\mathcal{L}_{\mathrm{m}})}{\delta g^{\mu\nu}}=g_{\mu\nu}\mathcal{L}_{\mathrm{m}}-2\frac{\partial \mathcal{L}_{\mathrm{m}}}{\partial g^{\mu\nu}}.
\end{align}
Einstein's field equations are represented as,
\begin{align} \label{Gmunu}
    G_{\mu\nu}=\kappa T_{\mu\nu},
\end{align}
where $G_{\mu \nu }=\mathcal{R}_{\mu \nu }-\frac{1}{2}g_{\mu \nu }\mathcal{R}$, is the Einstein tensor. 
\\
The ideal fluid form of $T_{\mu\nu}$ is given by
\begin{align}  \label{em}
T_{\mu\nu}=(\mathcal{E}+P)u_{\mu}u_{\nu}+P g_{\mu\nu},
\end{align}
where $\mathcal{E} $ and $P$ are the energy density and pressure respectively. $u_{\mu }$ is the four-velocity satisfying the conditions $u_{\mu }u^{\mu }=-1$, $\nabla _{\nu }u^{\mu }u_{\mu }=0$. 
\\
After adding the extra function ($f$), we get the modified total matter Lagrangian as $\mathcal{L}_{\mathrm{m}}^{tot}=\mathcal{L}_{\mathrm{m}}+f$, so as $T_{\mu\nu}$ from Eq. \eqref{tmunudef}. 
Accordingly, we will get modified Einstein equations. 
\\
Now, by arranging all the terms and considering $\mathcal{L}_m = P$ \cite{Faraoni, AkarsuNew, PhysRevD.94.044002, EMSG_NAlam} to construct a viable astrophysical/cosmological model, we will get 
\begin{eqnarray}
& & G_{\mu\nu}=\kappa \mathcal{E} \left[\left(1+\frac{P}{\mathcal{E}}\right)u_{\mu}u_{\nu}+\frac{P}{\mathcal{E}}g_{\mu\nu}\right] \nonumber \\
&  &+\alpha\kappa\mathcal{E}^2\left[2\left(1+\frac{4P}{\mathcal{E}}+\frac{3P^2}{\mathcal{E}^2}\right)u_{\mu}u_{\nu}+\left(1+\frac{3P^2}{\mathcal{E}^2}\right)g_{\mu\nu}\right].\nonumber\\
\label{fieldeq2}
\end{eqnarray}
Now we can restore the Einstein field equation by redefining the Eq.~\eqref{fieldeq2} as $G^{\mu\nu} = \kappa T^{\mu\nu}_{\mathrm{eff}}$, where $T^{\mu\nu}_{\mathrm{eff}} = (\mathcal{E}_{\mathrm{eff}}+P_{\mathrm{eff}})u^{\mu}u^{\nu} + P_{\mathrm{eff}}g^{\mu\nu}$, is the effective energy-momentum tensor.
For an ideal fluid, $\mathcal{E}_{\mathrm{eff}}$ is the effective energy density defined as
\begin{equation}
\label{Eeff}
    \mathcal{E}_{\mathrm{eff}} = \mathcal{E} + \alpha\mathcal{E}^2\Bigg(1+\frac{8P}{\mathcal{E}} + \frac{3P^2}{\mathcal{E}^2}\Bigg)\,,
\end{equation}
and $P_{\mathrm{eff}}$ is the effective pressure defined as
\begin{equation}
\label{Peff}
    P_{\mathrm{eff}} = P + \alpha\mathcal{E}^2\Bigg(1+\frac{3P^2}{\mathcal{E}^2}\Bigg)\,.
\end{equation}
As mentioned in \cite{EMSG_OAkarsu}, the parameter $|\alpha |\sim\mathcal{E} ^{-1}$, and we know that for NSs, $\mathcal{E}\sim 10^{37}\,\mathrm{erg\,cm^{-3}}$ \cite{shapiro}. Consequently, EMSG corrections are expected to become important for compact objects such as NSs when the order of $\alpha$ is approximately $|\alpha |\sim 10^{-37}\,\mathrm{erg^{-1}\,cm^{3}}$.
\subsection{Modified TOV Equations}
\label{sec:hydrostatics}
\begin{figure*}
    
    \includegraphics[width=0.495\textwidth]{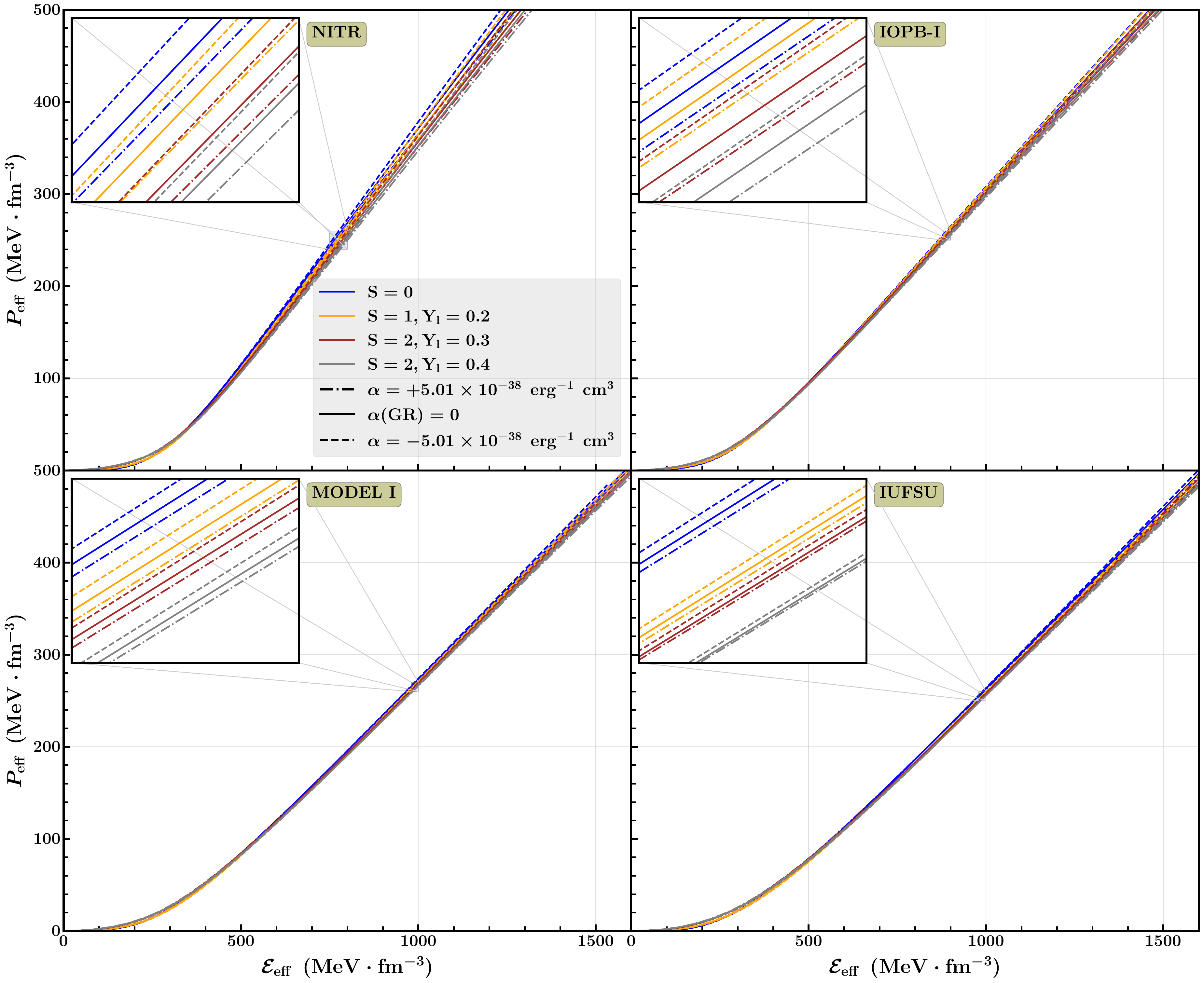}
    \includegraphics[width=0.5\textwidth]{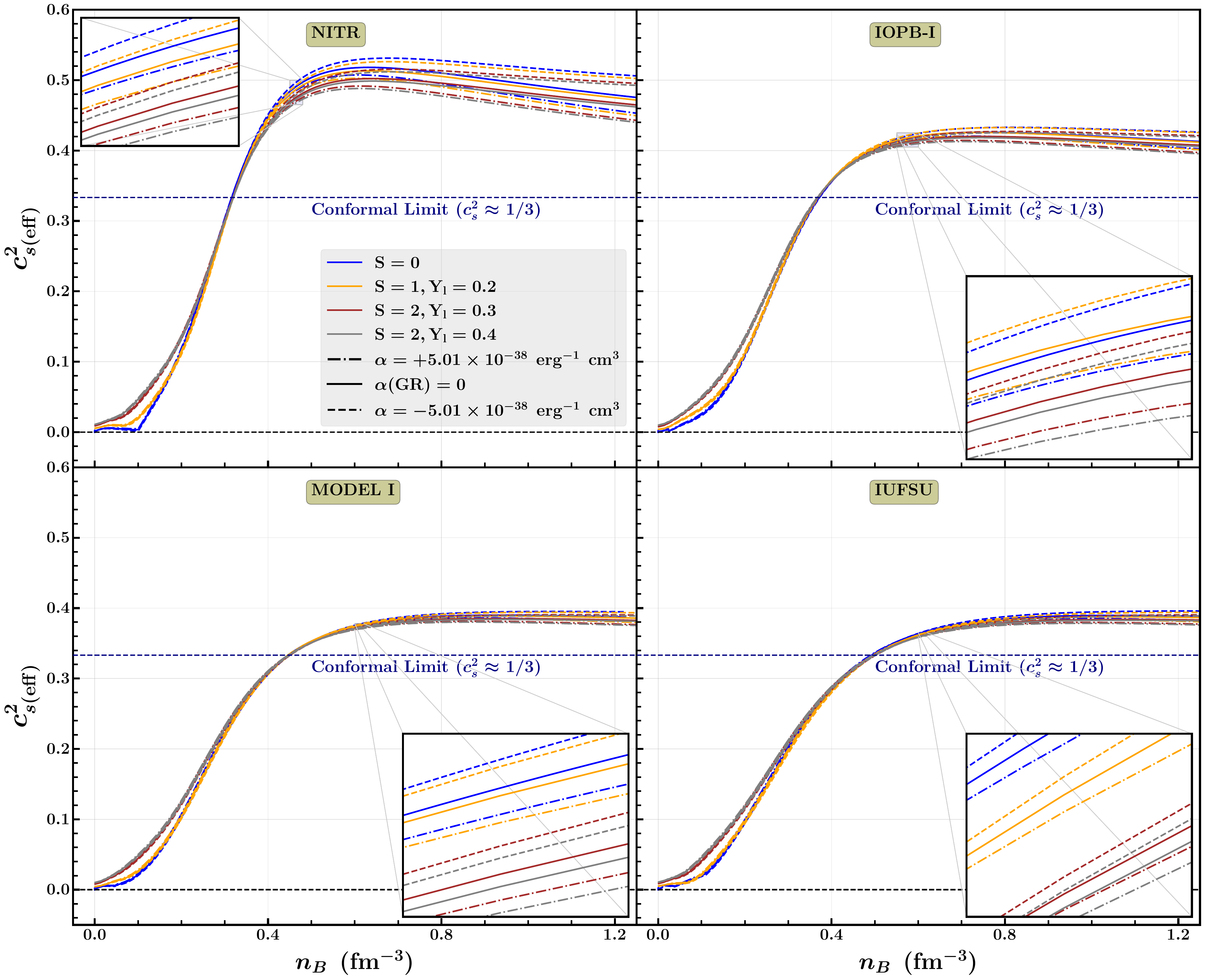}
    \caption{ \textit{Left:} Variation of effective pressure ($P_\mathrm{eff}$) with effective energy density ($\mathcal{E}_\mathrm{eff}$) for NITR, IOPB-I, MODEL I and IUFSU EOSs, represented by different colours. \textit{Right:} The effective squared sound speeds ($c^2_{s(\mathrm{eff})}$) are presented as functions of baryon number density ($n_B$). In both panels, the zoom plots refer to the clear visibility of the effect of $S$, $Y_l$ and $\alpha$.}
    \label{EOS_SS}
\end{figure*}

The line element describes the spacetime geometry inside a static, spherically symmetric star \cite{Wald:1984rg, Schwarzschild:1916uq}:
\begin{equation}  
\label{eqn:metric}
\mathrm{d} s^2 = -e^{2\nu\left(r\right)}\mathrm{d} t^2 +e^{2\lambda\left(r\right)}\mathrm{d} r^2+r^2\mathrm{d}\theta^2+r^2\sin^2\theta \, \mathrm{d}\phi^2
\end{equation}
where $\nu(r)$ and $\lambda(r) $ are the metric functions. Now, according to Eq. \eqref{fieldeq2}, solving the modified Einstein equations in the EMSG framework, we get the modified Tolman–Oppenheimer–Volkoff (TOV) equations \cite{AkarsuNew, EMSG_NAlam, Ghosh2025, ghosh2025curv}, as,
\begin{align}
\frac{\mathrm{d} m}{\mathrm{d} r}=4\pi r^2 \mathcal{E} \left[1+\alpha\mathcal{E} \left(1+\frac{8P}{\mathcal{E}}+\frac{3P^2}{\mathcal{E}^2 }\right)\right],  \label{TOV1}
\end{align}
\begin{align}
\frac{\mathrm{d} P}{\mathrm{d} r}&= -\frac{m\mathcal{E} }{r^2}\left(1+\frac{P}{\mathcal{E}}\right) \left( 1-\frac{2m}{r}\right)^{-1}  \notag \\
&\times \left[ 1+\frac{4\pi r^3 P}{m }+\alpha \frac{4\pi r^3\mathcal{E}^2}{m}\left(1+\frac{3P^2}{\mathcal{E}^2}\right)\right] \notag \\
&\times \left[1+2\alpha\mathcal{E}\left(1+\frac{3P}{\mathcal{E}} \right)\right] \left[1 + 2\alpha\mathcal{E} \left(c_s^{-2}+\frac{3P}{\mathcal{E}}\right)\right]^{-1}.  \label{TOV2}
\end{align}
We solved the modified TOV equations by selecting an EOS governing the radial pressure ($P$) and adopting the EMSG model for $\alpha$. It becomes feasible to solve Eqs. \eqref{TOV1}-\eqref{TOV2} numerically, by integrating from $r = 0$ where $m(r = 0) = 0$ and $P (r = 0) = P_c $ (the central pressure) upto the stellar surface $r = R$ where $m(r = R) = M$ and $P (r = R)=0 $. This integration was performed for four temperature-dependent EOSs to get the PNS mass-radius profile.
\\
These four EOSs with the variation of $S$, $Y_l$ and $\alpha$ are shown in the left panel of Fig. \ref{EOS_SS}. In these figures, different colours represented the different combinations of $S$ and $Y_l$, and different linestyles are for different values of $\alpha$. Also, to validate the causality condition, we have shown the variation of sound speed squared, defined as $c_{s}^2 \equiv \frac{\mathrm{d} P}{\mathrm{d} \mathcal{E}}$ (in units of $c^2$; $c$ is the speed of light), with baryon number density ($n_B$), in the right panel of Fig. \ref{EOS_SS} (We have represented the effective sound speed squared $c_{s (\mathrm{eff})}^2 \equiv \frac{\mathrm{d} P_\mathrm{eff}}{\mathrm{d} \mathcal{E}_\mathrm{eff}}$, after getting the effective pressure as $P_\mathrm{eff}$ and effective energy density as  $\mathcal{E}_\mathrm{eff}$, from Eqs. \eqref{Eeff} and \eqref{Peff}). The causality limit is $0 \le c_{s}^2 \le 1$, and our calculated effective sound speed values are between $0$ to $0.6$. Accordingly, the causality condition is verified for all $S$, $Y_l$, and $\alpha$ variations.
\subsection{$f$-mode Oscillations}
\label{fmode}
NSs or PNSs oscillate when they go through external or internal perturbations, and they emit different mode frequencies \cite{PhysRevD.66.104002}. In this study, we solve the fundamental ($f$)-mode nonradial oscillations of spherically symmetric PNSs by using the Cowling approximation \cite{10.1093/mnras/101.8.367}, which ignores the metric perturbation by assuming the spacetime is frozen. These modes have been extensively studied in the context of compact objects like NSs or PNSs to probe their internal structure and dynamic behaviour \cite{Ghosh2025, GHOSH_FTT}.
\\
The Lagrangian displacement vector of the fluid is given by \\
\begin{equation}
\xi^{i}=\frac{1}{r^2}\Big(e^{-\lambda (r)}W (r),-V
(r)\partial_{\theta},
     -\frac{V(r)}{ \sin^{2}{\theta}}\  \partial _{\phi}\Big)
e^{i\omega t}Y_{lm}(\theta,\phi) \textcolor{red}{,}
\end{equation}
where $Y_{lm}(\theta,\phi)$ represents the spherical harmonics, $\omega$ is the frequency. To determine $\omega$, we have to solve the following system of ordinary differential equations \cite{PhysRevD.83.024014}, which are modified by the effective energy density and pressure in EMSG, represented as, 
\begin{eqnarray}
\frac{d W(r)}{dr}&=&\frac{d {\cal E}_{\mathrm{eff}}}{dP_{\mathrm{eff}}}\left[\omega^2r^2e^{\lambda
(r)-2\nu (r)}V(r)
+\frac{d \nu(r)}{dr} W (r)\right] \nonumber \\
&&
-l(l+1)e^{\lambda (r)}V (r) \nonumber \\
\frac{d V(r)}{dr} &=& 2\frac{d\nu (r)}{dr} V
(r)-\frac{1}{r^2}e^{\lambda (r)}W (r),
\label{eqn:cowling}
\end{eqnarray}
where $\nu(r)$ and $\lambda(r)$ represent metric functions, with the fixed background metric as mentioned in Eq. \eqref{eqn:metric}.
\\
In the close vicinity of the origin, the solution to Eq. (\ref{eqn:cowling}) exhibits the following behaviour:
\begin{equation}
     W (r)=Br^{l+1}, \ V (r)=-\frac{B}{l} r^l,
\label{eq:bc1}
\end{equation}
where $B$ is an arbitrary constant. To ensure that the perturbation pressure becomes zero at the outer boundary of the star's surface, we need to apply the following additional boundary condition,
\begin{equation}
     \omega^2 e^{\lambda (R)-2\nu (R)}V (R)+\frac{1}{R^2}\frac{d\nu
(r)}{dr}\Big|_{r=R}W (R)=0.
\label{eq:bc2}
\end{equation}
Utilizing the boundary conditions outlined in Eq. (\ref{eq:bc1}) and Eq. (\ref{eq:bc2}), we can successfully solve Eq. (\ref{eqn:cowling}) and determine the eigenfrequencies of the PNSs.
\subsection{Gravitational binding energy}
The gravitational binding energy ($B$) is defined as the difference between the gravitational mass ($M$) and  baryonic mass ($M_B$) of the NS and PNS;  $B=M-M_B$, where $M$ is calculated \cite{Glendenning, Xiao_2015}, and modified by effective energy density in EMSG as
\begin{equation}
M=\int_{0}^{R} dr \  4\pi r^2 {\cal E}_{\mathrm{eff}}(r),
\end{equation}
and $M_B=Nm_b$, where $m_b$ is the mass of baryons (931.5 MeV) and $N$ is the number of baryons calculated by integrating over the whole volume in the Schwarzchild limit as 
\begin{eqnarray}
N=\int_{0}^{R} dr \ 4\pi r^2 \Big[1-\frac{2m(r)}{r}\Big]^{-1/2}.
\end{eqnarray}
In our work binding energy $B$ originally corresponds to $B/M$, which is more convenient for comparison purposes.
$B/M$ is positive for an unstable system and negative for a stable system \citep{Serot_1986}. For the whole NS, the $B$ is negative \citep{Patra_1992,Satpathy_2004,Kaur_2020}.
\section{Results and Discussion} 
\label{sec:R&D}
In this section, we present the numerical results for four different EOSs, such as NITR, IOPB-I, IUFSU and MODEL I, focusing on how the EMSG parameter $\alpha$ affects the macroscopic properties of proto-neutron stars, with the variation of entropy at $S=0$ (cold NS) and $S=1,2$ (PNS with trapped
neutrinos at $Y_l=0.2-0.4$).
\begin{figure*}
    
    \includegraphics[width=0.52\textwidth]{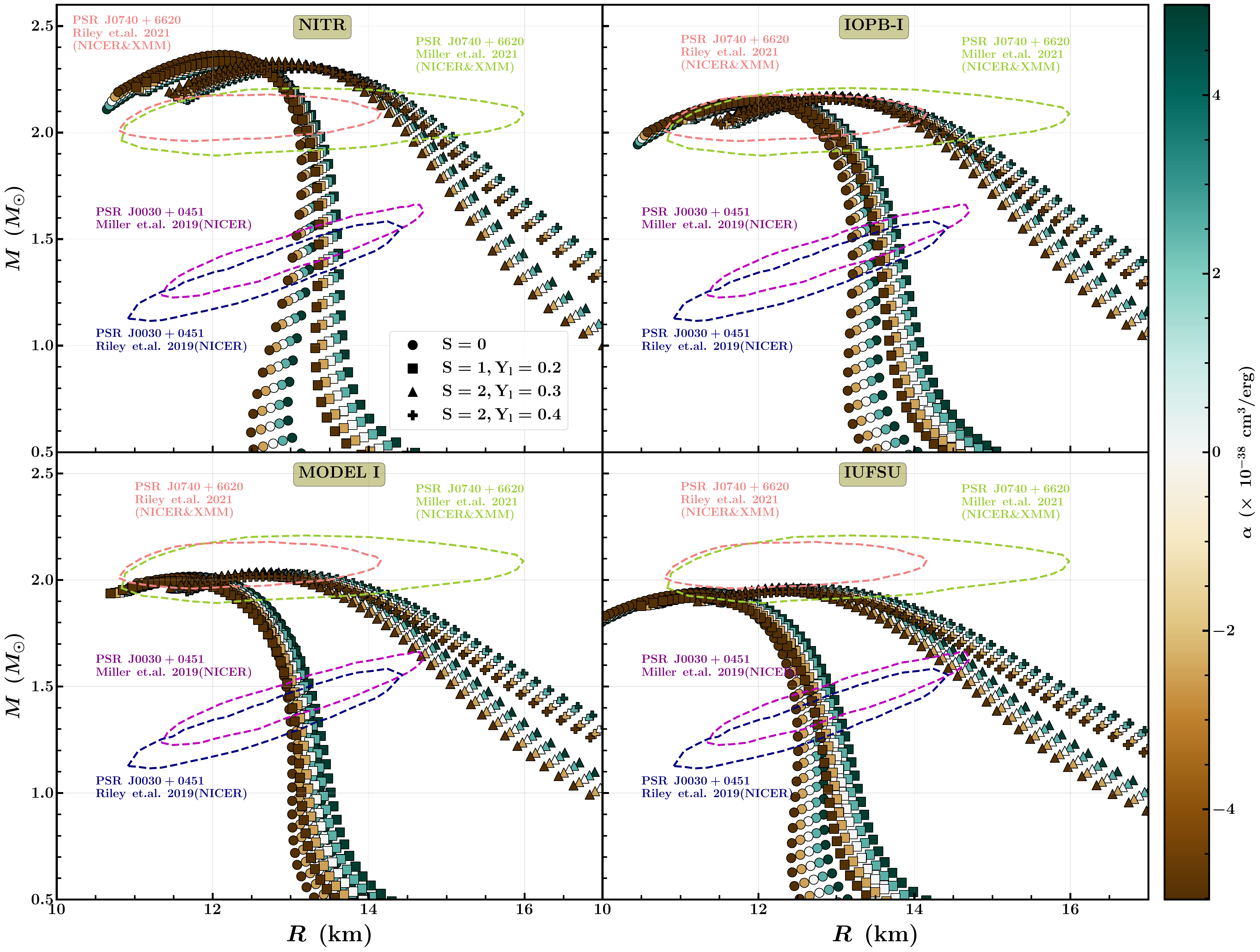}
    \includegraphics[width=0.475\textwidth]{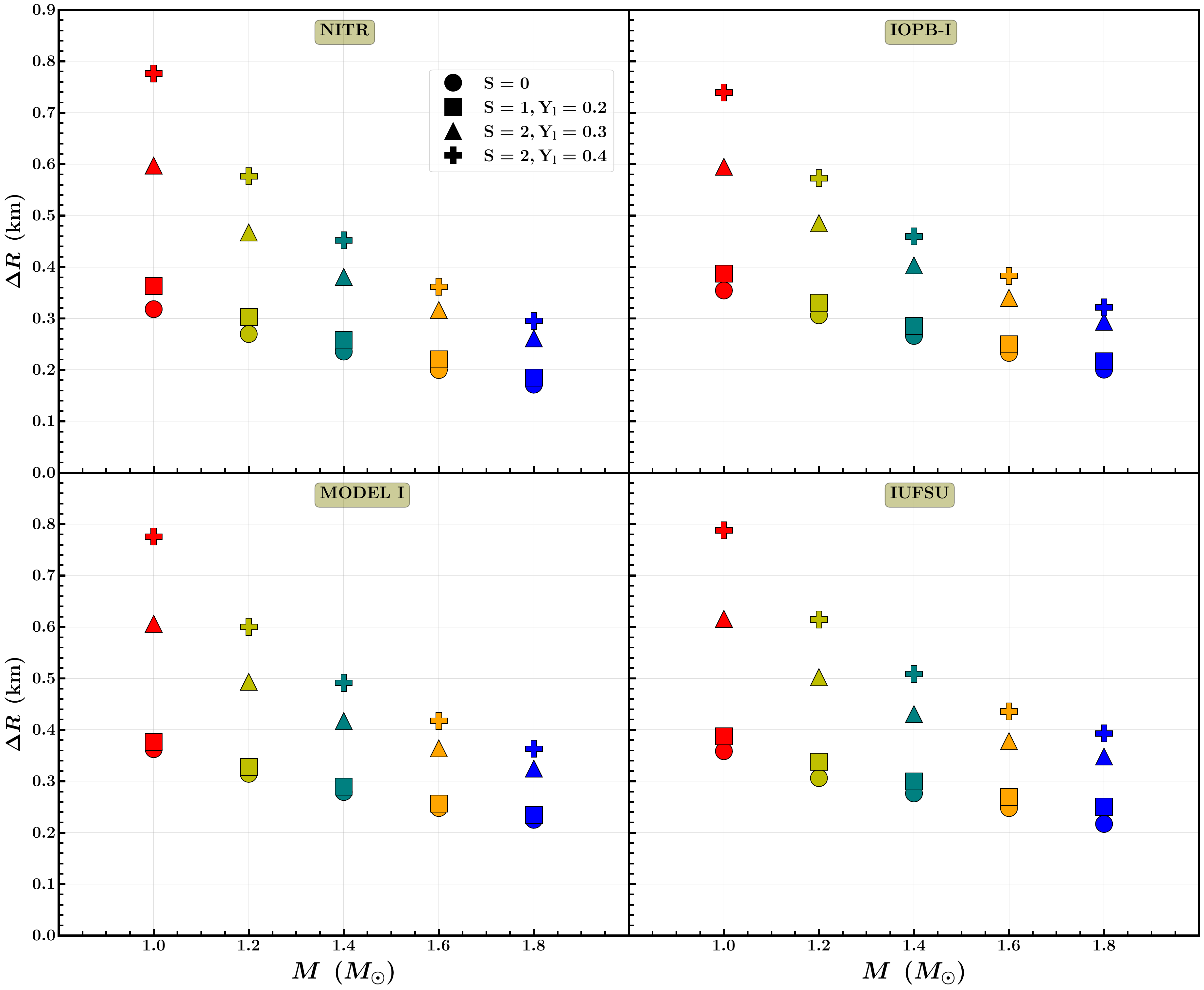}
    \caption{ \textit{Left:} The Mass-radius relation of PNSs, for differnet $S$ and $Y_l$ values. The astrophysical observable constraints on mass and radius from PSR J0740+6620 \cite{Miller_2021, Riley_2021}, and NICER data for PSR J0030+0451 \cite{Miller_2019, Riley_2019} are represented by colored regions. In the colour bar, the variation of $\alpha$ has been shown. \textit{Right:} Variation of the difference in $R$ for maximum and minimum values of $\alpha$ with fixed mass represented with different coloured markers.}
    \label{MR}
\end{figure*}
In the left panel of Fig. \ref{MR}, we display the mass-radius relation for varying the entropy and lepton fraction. Each band represents the different entropies and lepton fractions represented by different markers, and the variation in the $\alpha$ parameter is shown in the colour bar; each colour corresponds to a different $\alpha$ value. Increasing entropy and lepton fraction means we are going back toward PNS, where the thermal pressure is higher than NS. That's why the PNS with higher entropy, i.e. with higher thermal pressure, leads to a larger radius at a given mass and the maximum mass also slightly increases. More interestingly, higher entropy flattens the mass-radius curve near the stability region (Maximum mass), indicating PNSs can support a broad range of radii near the stability limit. However, for fixed entropy, the larger trapped lepton fraction tends to decrease the maximum mass. These effects of $S$ and $Y_l$ are consistent with prior PNS studies. Our results remain compatible with the observational constraints on NS masses/radii. The $S=0$ (NS) and $S=1$ curves lie within the region of PSR J0740+6620 ($M\approx2.1\,M_\odot$, $R\sim13$~km) and PSR J0030+0451 ($1.4\,M_\odot$, $R\sim12$–13~km). However, for $S=2$ the radius increases more, so the curves corresponding to $S=2$ at maximum mass region only lie within the region of PSR J0740+6620. 
\begin{figure*}
    
    \includegraphics[width=0.52\textwidth]{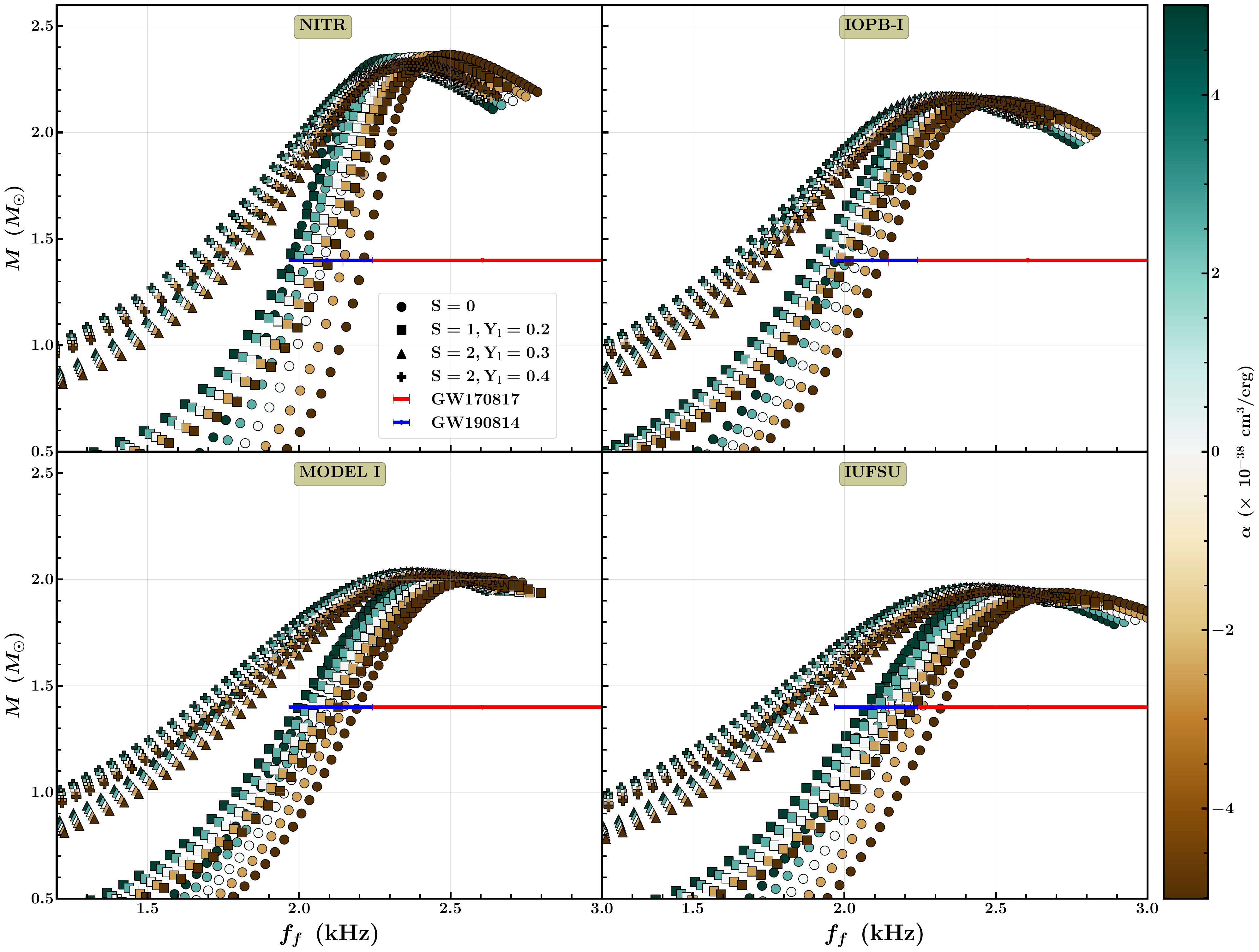}
    \includegraphics[width=0.475\textwidth]{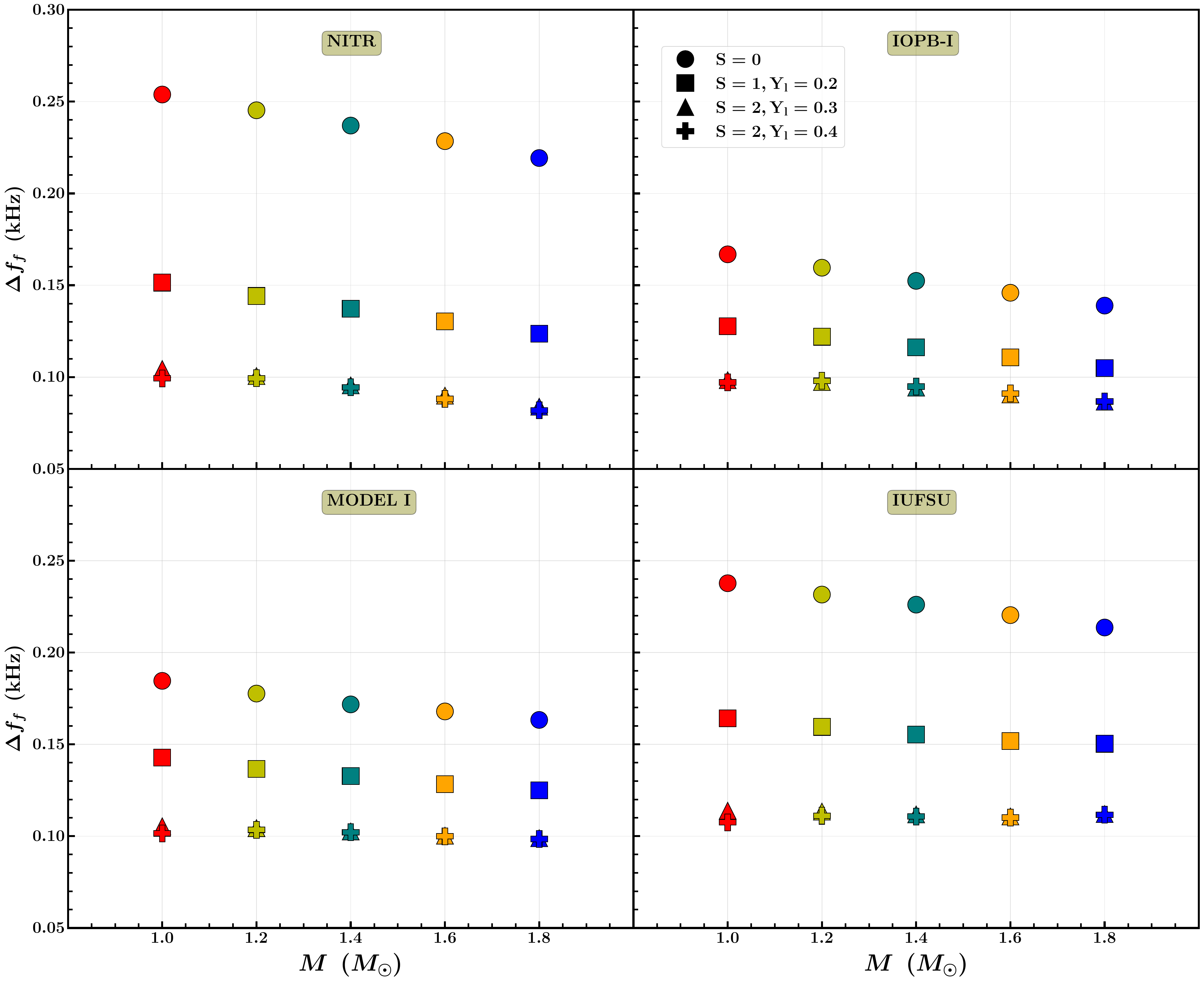}
    \caption{ \textit{Left:} The variation of $f$-mode frequency ($f_f$) with the mass ($M$) of PNSs, for differnet $S$ and $Y_l$ values. The error bars represent the observational constraints from GW170817 \cite{GW170817} and GW190814 events \cite{GW190814}. In the colour bar, the variation of $\alpha$ has been shown. \textit{Right:} Variation of the difference in $f_f$ for maximum and minimum values of $\alpha$ with fixed mass represented with different coloured markers.}
    \label{fM}
\end{figure*}
\begin{figure*}
    
    \includegraphics[width=0.52\textwidth]{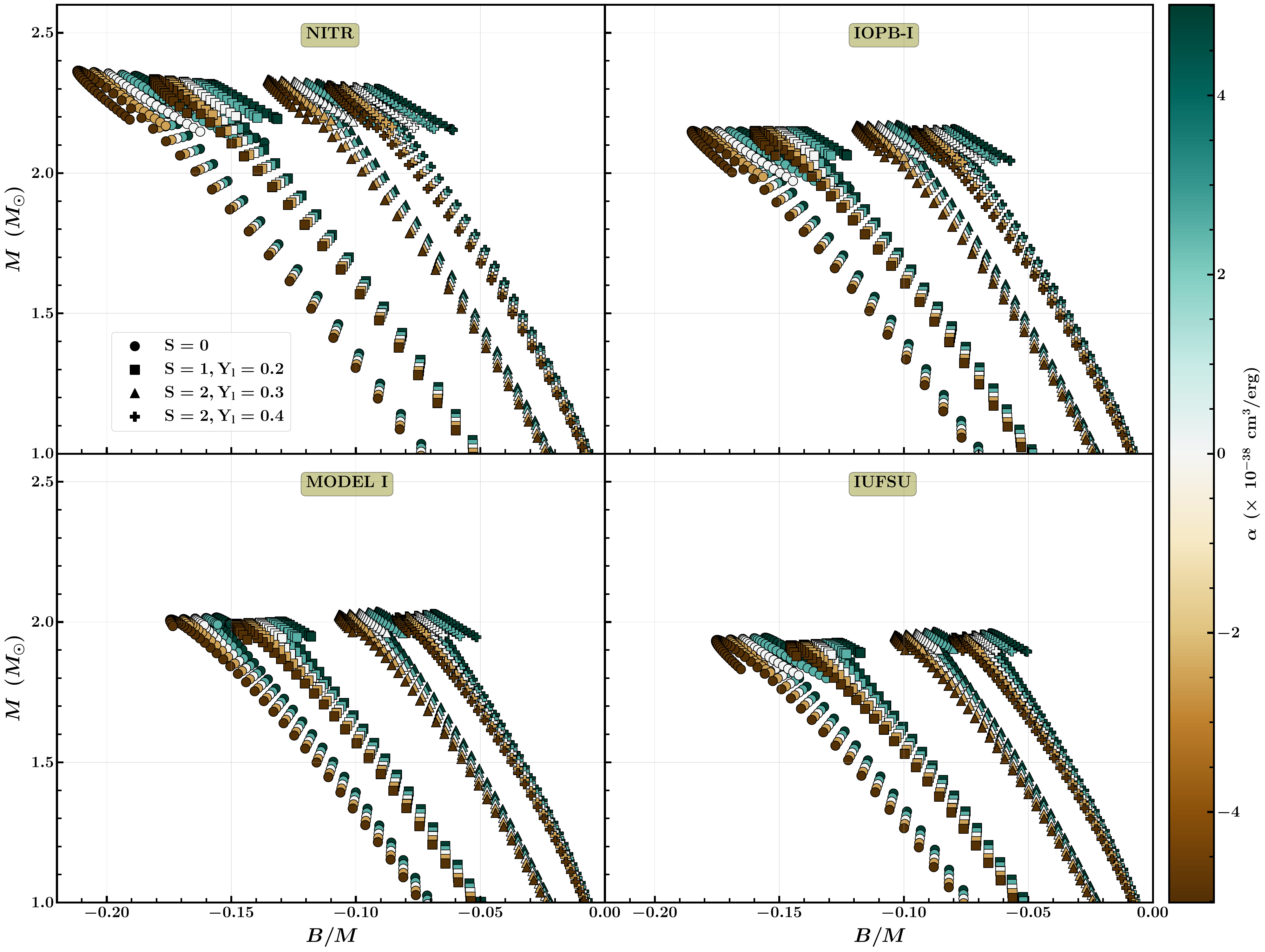}
    \includegraphics[width=0.475\textwidth]{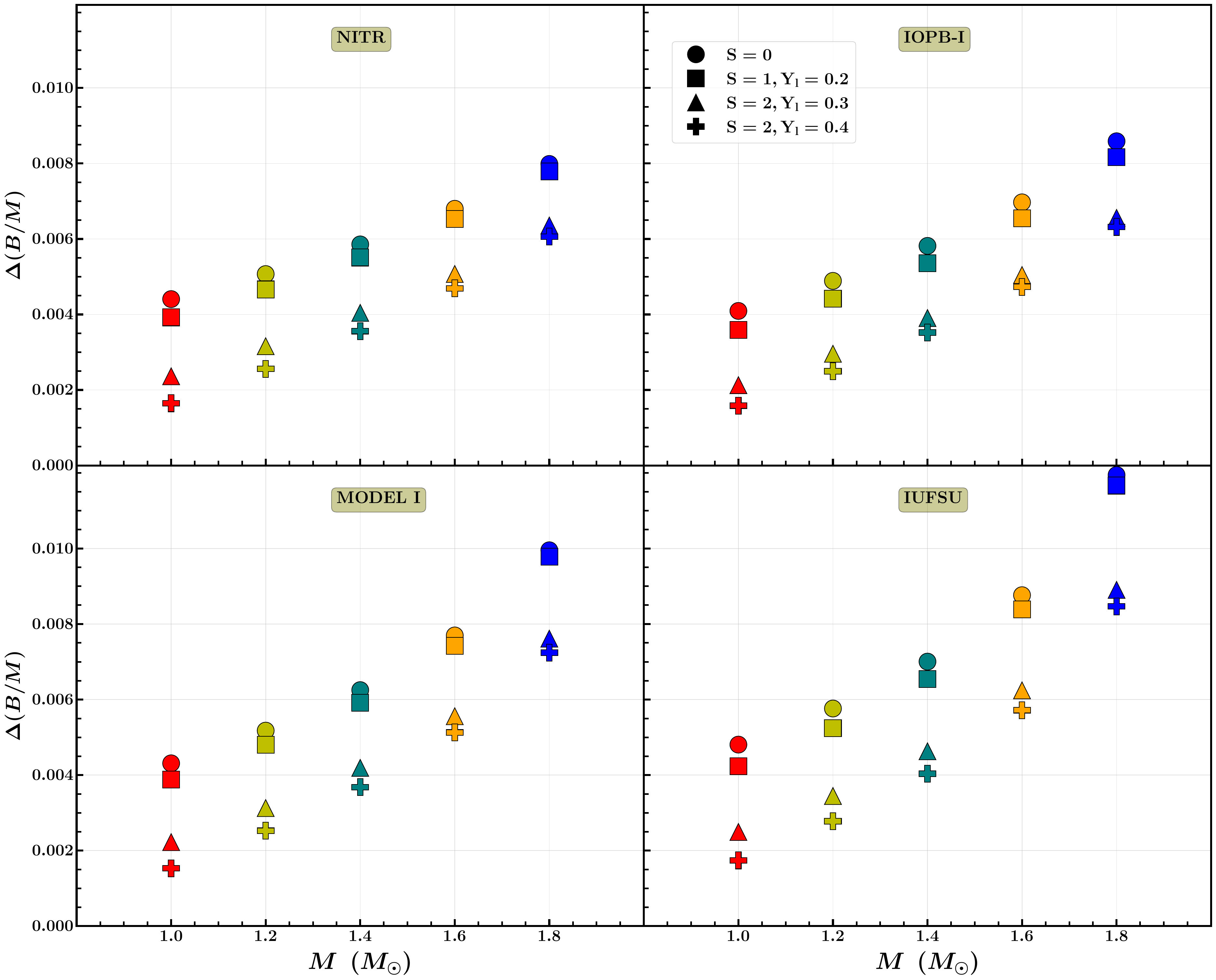}
    
    \caption{ \textit{Left:} The variation of gravitational binding energy per mass ($B/M$) with the mass ($M$) of PNSs, for differnet $S$ and $Y_l$ values. In the colour bar, the variation of $\alpha$ has been shown. \textit{Right:} Variation of the difference in $B/M$ for maximum and minimum values of $\alpha$ with fixed mass represented with different coloured markers.}
    \label{BM}
\end{figure*}
\\ The effect of EMSG is also illustrated in this figure. Along the stable branch of each curve (configurations below the turning point, i.e., below the maximum-mass configuration), positive $\alpha$ values produce moderately higher masses and slightly larger radii at fixed central density. In contrast, the negative $\alpha$ shifts the models toward marginally lower masses and smaller radii.  
This behaviour reflects the fact that, for positive $\alpha$, the effective pressure term in EMSG effectively stiffens the EOS at low and intermediate densities, whereas negative $\alpha$ has the opposite effect.  
Near the turning point, positive $\alpha$ leads to a small decrease in maximum mass ($M$) accompanied by a modest increase in the corresponding radius, whereas negative $\alpha$ slightly raises $M$ and contracts the star. Since the maximum mass configuration occurs at different central densities for different $\alpha$ values so their radius corresponding to maximum mass also differs for each curve.  
Overall, the maximum masses remain almost unchanged, but the radii vary noticeably with $\alpha$.  
Hence, relative to GR, the EMSG framework effectively stiffens the EOS at lower densities and softens it at higher densities for positive $\alpha$, with the inverse pattern for negative $\alpha$.
\\
The right panel of Fig. \ref{MR} represents the effect of $\alpha$ for NSs ($S=0$) and PNSs ($S=1,2$). For that, we calculate the width of the $M-R$ curve, i.e. the difference in the radius for maximum and minimum values of $\alpha$ for fixed mass. We take five different masses and calculate the $\Delta R$ for different $S$ and $Y_l$ values. We find that PNSs with higher $S$ and $Y_l$ ($S=2, Y_l=0.4$) have the more $\Delta R$ compared to the PNSs with lower $S$ and $Y_l$ and NSs ($S=0$). Therefore, we can say that the effect of the EMSG coupling parameter $\alpha$ on radius is more pronounced for PNSs with higher $S$ and $Y_l$ values.

The $f$-mode frequency plays a crucial role in PNSs to understand the oscillation behaviour and their interior properties. In the left panel of Fig. \ref{fM}, we show the variation of $f$-mode frequency with the mass ($M$) of the PNSs, for varying $S$ and $Y_l$, with the different values of $\alpha$ shown in the colour bar. It is clearly visible that below the maximum mass point, the frequency corresponding to higher $S$ and $Y_l$ is less than that corresponding to lower $S$ and $Y_l$ values at fixed mass. Also, at maximum masses, our results show a similar nature to the previous findings on PNSs. The effect of the EMSG coupling parameter $\alpha$ on the frequency is clearly visible. The negative values of $\alpha$ make the EOS softer (as shown in Fig. \ref{MR}), which means a less compact PNS, and we get a higher frequency. The opposite thing happens for positive $\alpha$. Now, to understand the effect of the coupling parameter $\alpha$ on frequency curves for different values of $S$ and $Y_l$, we plot the width of the frequency curve, i.e. the difference of the frequencies at maximum and minimum values of $\alpha$ at different fixed masses in the right panel of Fig. \ref{fM}. Where we find that, for the $S=0$ (NSs) case, $\Delta f_f$ is large compared to $S=1,2$ and $Y_l$ values. It is obvious from the right panel of Fig. \ref{MR} that the $\Delta R$ is large for $S=2, Y_l=0.4$, so $\Delta f_f$ is small for the same.

In the left panel of Fig. \ref{BM}, we present the variation of the gravitational binding energy per mass ($B/M$) with the mass of the PNSs. We can clearly see that, for all variations of $ S$, $ Y_l$, and $\alpha$, the values of $B/M$ are negative across the entire mass range, indicating that the PNSs will remain a bound system for all variations. However, if we focus on increasing the values of $ S$ and $ Y_l$, the PNSs are going towards the positive values of $B/M$. So, we can say that the NSs are a more bound system than the PNSs, where there is a tendency of the PNSs to be in an unbound system, but they remain in a bound system for this particular range of $ S$ and $ Y_l$. Also, the positive (negative) values of $\alpha$ shift the curves towards the positive (negative) values of $B/M$. The effect of $\alpha$ is less near the low mass region and more near the high mass region.
To understand the effect of the coupling parameter $\alpha$ on $B/M$ curves for different values of $S$ and $Y_l$, we plot the width of the $B/M$ curve, i.e. the difference of the $B/M$ at maximum and minimum values of $\alpha$ at different fixed masses in the right panel of Fig. \ref{BM}. Where we find that, for the $S=0$ (NSs) case, $\Delta B/M$ is more compared to $S=1,2$ and $Y_l$ values.
\section{Universal Relations}
\label{sec:UR}
\begin{table*}
\caption{The fitting coefficients are listed  for  $f_fM_{1.4}$-$C$ relation with $\alpha (10^{-38}) = +5.01,  0,  -5.01$. The reduced chi-squared ($\chi_r^2$) is also given.}
\centering
\setlength{\tabcolsep}{4.5pt}
\renewcommand{\arraystretch}{1.5}
\scalebox{1}{
    \begin{tabular}{cccccccccccccccc}
        \hline \hline
        & \multicolumn{3}{c}{$S=0$} && \multicolumn{3}{c}{$\mathrm{S=1,Y_l=0.2}$}&& \multicolumn{3}{c}{$\mathrm{S=2,Y_l=0.3}$} && \multicolumn{3}{c}{$\mathrm{S=2,Y_l=0.4}$} \\
            \cline {2-4} \cline {6-8} \cline {10-12} \cline {14-16}
        $\alpha (10^{-38}) =$ & $+5.01$  & 0  & $-5.01$ && $+5.01$ &  0  &  $-5.01$ && $+5.01$ &  0  &  $-5.01$ && $+5.01$ &  0  &  $-5.01$ \\
        \hline
        $a_0\left(10^{-2}\right)=$ &2.200  &-2.348   &-6.740   &&-6.781  &-8.185  &-9.410   &&-18.13   &-20.82    &-23.59    &&-36.88    &-41.26    &-46.40    \\
        $a_1=$ &6.999  &10.01   &13.14   &&7.156  &8.064  &8.922   &&9.583   &10.68    &11.84    &&14.41    &15.98    &17.85    \\
        $a_2\left(10^{+1}\right)=$ &8.018  &2.974   &-2.317   &&8.590  &7.280  &6.064   &&8.138   &6.795    &5.325    &&3.644    &1.766    &-0.567    \\
        $a_3\left(10^{+2}\right)=$ &-5.026  &-0.806   &3.742   &&-4.917  &-3.752  &-2.627   &&-5.000   &-4.040    &-2.971    &&-2.622    &-1.380    &0.185    \\
        $a_4\left(10^{+3}\right)=$ &1.856  &0.286   &-1.441   &&1.597  &1.142  &0.689   &&1.615   &1.287    &0.917    &&0.955    &0.556    &0.048    \\
        $a_5\left(10^{+3}\right)=$ &-2.783  &-0.664   &1.702   &&-2.202  &-1.569  &-0.927   &&-2.217   &-1.797    &-1.323    &&-1.502    &-1.014    &-0.390    \\
        $\chi_r^2\left(10^{-4}\right)=$ &5.089  &1.375   &4.492   &&1.128  &1.432  &2.669   &&1.343   &1.582    &1.868    &&1.291    &1.441    &1.615    \\
        \hline \hline
    \end{tabular}}
    \label{f-c}
\end{table*}
\begin{table*}
\caption{The fitting coefficients are listed  for  $f_fM_{1.4}$-$\Lambda$ relation with $\alpha (10^{-38}) = +5.01,  0,  -5.01$. The reduced chi-squared ($\chi_r^2$) is also given.}
\centering
\setlength{\tabcolsep}{4.5pt}
\renewcommand{\arraystretch}{1.5}
\scalebox{1}{
    \begin{tabular}{cccccccccccccccc}
        \hline \hline
        & \multicolumn{3}{c}{$S=0$} && \multicolumn{3}{c}{$\mathrm{S=1,Y_l=0.2}$}&& \multicolumn{3}{c}{$\mathrm{S=2,Y_l=0.3}$} && \multicolumn{3}{c}{$\mathrm{S=2,Y_l=0.4}$} \\
            \cline {2-4} \cline {6-8} \cline {10-12} \cline {14-16}
        $\alpha (10^{-38}) =$ & $+5.01$  & 0  & $-5.01$ && $+5.01$ &  0  &  $-5.01$ && $+5.01$ &  0  &  $-5.01$ && $+5.01$ &  0  &  $-5.01$ \\
        \hline
        $b_0=$ & 4.148 &  4.415 &  4.748 &&4.338&4.454&4.577 &&4.555&4.614&4.668 &&4.765&4.783&4.801\\
        $b_1\left(10^{-1}\right)=$ & 2.155  & -0.503  & -3.992 &&-0.312&-0.964&-1.646 &&-6.124&-6.142&-6.000  &&-12.16&-11.15&-10.09\\
        $b_2\left(10^{-1}\right)=$ & -7.563  & -5.930  & -3.825 &&-6.313&-5.972&-5.619 &&-1.280&-1.432&-1.713 &&4.811&3.622&2.399\\
        $b_3\left(10^{-1}\right)=$ & 1.960  & 1.471  & 0.873 &&1.650&1.549&1.446 &&-0.369&-0.278&-0.140 &&-3.230&-2.648&-2.061\\
        $b_4\left(10^{-2}\right)=$ & -2.010  & -1.342  & -0.574 &&-1.652&-1.521&-1.390 &&2.057&1.831&1.527 &&8.263&6.944&5.638\\
        $b_5\left(10^{-4}\right)=$ & 7.464  & 4.084  & 0.442 &&5.945&5.335&4.743 &&-19.38&-17.43&-14.99 &&-69.81&-58.65&-47.77\\
        $\chi_r^2\left(10^{-4}\right)=$ & 7.222  & 10.430  & 27.390 &&5.686&6.787&9.419  &&6.917&6.855&6.840 &&7.209&6.968&6.726\\
        \hline \hline
    \end{tabular}}
    \label{f-Love}
\end{table*}
\begin{table*}
\caption{The fitting coefficients are listed  for  $|B|/M$-$C$ relation with $\alpha (10^{-38}) = +5.01,  0,  -5.01$. The reduced chi-squared ($\chi_r^2$) is also given.}
\centering
\setlength{\tabcolsep}{4.5pt}
\renewcommand{\arraystretch}{1.5}
\scalebox{1}{
    \begin{tabular}{cccccccccccccccc}
        \hline \hline
        & \multicolumn{3}{c}{$S=0$} && \multicolumn{3}{c}{$\mathrm{S=1,Y_l=0.2}$}&& \multicolumn{3}{c}{$\mathrm{S=2,Y_l=0.3}$} && \multicolumn{3}{c}{$\mathrm{S=2,Y_l=0.4}$} \\
            \cline {2-4} \cline {6-8} \cline {10-12} \cline {14-16}
        $\alpha (10^{-38}) =$ & $+5.01$  & 0  & $-5.01$ && $+5.01$ &  0  &  $-5.01$ && $+5.01$ &  0  &  $-5.01$ && $+5.01$ &  0  &  $-5.01$ \\
        \hline
        $c_0\left(10^{-2}\right)=$ &-1.254  &-0.716   &-0.309   &&-2.056  &-1.700  &-1.332   &&-7.013   &-4.918    &-3.286    &&-12.15    &-6.855    &-3.152    \\
        $c_1=$ &1.605  &1.277   &1.020   &&1.134  &0.936  &0.730   &&2.843   &1.961    &1.254    &&4.313    &2.343    &0.922    \\
        $c_2\left(10^{+1}\right)=$ &-2.211  &-1.542   &-0.998   &&-1.300  &-0.917  &-0.518   &&-3.977   &-2.613    &-1.492    &&-5.978    &-3.207    &-1.159    \\
        $c_3\left(10^{+2}\right)=$ &1.956  &1.404   &0.950   &&1.185  &0.882  &0.567   &&3.041   &2.077    &1.272    &&4.300    &2.466    &1.085    \\
        $c_4\left(10^{+2}\right)=$ &-7.139  &-5.172   &-3.548   &&-4.390  &-3.339  &-2.239   &&-10.48   &-7.319    &-4.647    &&-14.32    &-8.563    &-4.166    \\
        $c_5\left(10^{+2}\right)=$ &9.010 &6.564   &4.562   &&5.579  &4.317  &2.989   &&13.13   &9.313    &6.071    &&17.69    &10.87    &5.637    \\
        $\chi_r^2\left(10^{-5}\right)=$ &12.66  &7.990   &4.292   &&5.079  &3.659  &2.378   &&5.969   &4.262    &2.761    &&6.375    &4.536    &2.931    \\
        \hline \hline
    \end{tabular}}
    \label{B-C}
\end{table*}
\begin{table*}
\caption{The fitting coefficients are listed  for  $|B|/M$-$\Lambda$ relation with $\alpha (10^{-38}) = +5.01,  0,  -5.01$. The reduced chi-squared ($\chi_r^2$) is also given.}
\centering
\setlength{\tabcolsep}{4.5pt}
\renewcommand{\arraystretch}{1.5}
\scalebox{1}{
    \begin{tabular}{cccccccccccccccc}
        \hline \hline
        & \multicolumn{3}{c}{$S=0$} && \multicolumn{3}{c}{$\mathrm{S=1,Y_l=0.2}$}&& \multicolumn{3}{c}{$\mathrm{S=2,Y_l=0.3}$} && \multicolumn{3}{c}{$\mathrm{S=2,Y_l=0.4}$} \\
            \cline {2-4} \cline {6-8} \cline {10-12} \cline {14-16}
        $\alpha (10^{-38}) =$ & $+5.01$  & 0  & $-5.01$ && $+5.01$ &  0  &  $-5.01$ && $+5.01$ &  0  &  $-5.01$ && $+5.01$ &  0  &  $-5.01$ \\
        \hline
        $d_0\left(10^{-1}\right)=$ &0.863  &1.411   &1.972   &&1.042  &1.444  &1.849   &&0.807   &1.205    &1.602    &&0.975    &1.284    &1.595    \\
        $d_1\left(10^{-1}\right)=$ &1.549  &0.879   &0.212   &&0.850  &0.422  &0.008   &&0.149   &-0.250    &-0.622    &&-1.039    &-1.180    &-1.305    \\
        $d_2\left(10^{-2}\right)=$ &-10.42  &-7.051   &-3.792   &&-6.382  &-4.490  &-2.766   &&1.417   &2.804    &3.912    &&13.96    &12.60    &11.10    \\
        $d_3\left(10^{-3}\right)=$ &24.81  &16.61   &8.884   &&14.39  &10.32  &6.847   &&-19.37   &- 20.55   &-20.70    &&-78.88    &-66.67    &-54.22    \\
        $d_4\left(10^{-4}\right)=$ &-26.54  &-16.93   &-8.202   &&-13.85  &-9.664  &-6.375   &&51.83   &49.19    &44.94    &&183.2    &150.2    &117.6    \\
        $d_5\left(10^{-5}\right)=$ &10.77  &6.445   &2.652   &&4.827  &3.209  &2.065   &&-42.34   &-38.27    &-33.33    &&-151.1    &-121.1    &-92.35    \\
        $\chi_r^2\left(10^{-5}\right)=$ &13.35  &9.960   &6.950   &&6.010  &4.950  &3.878   &&6.452   &5.233    &4.030    &&6.669    &5.387    &4.140    \\
        \hline \hline
    \end{tabular}}
    \label{B-Love}
\end{table*}
The primary objective of URs is to investigate the properties of NSs that are challenging to measure through observational means. Numerous URs have been proposed to estimate the properties of NS, with most studies concentrating on isotropic cases \cite{PhysRevD.101.124006, Breu_2016, PhysRevD.91.044034, Yagi:2013bca, Kent_yagi_2013, staykov2016}. In our recent research \cite{Mohanty_2024}, we examined the URs for anisotropic NSs, and in \cite{Ghosh2025}, we have analyzed URs for NSs within the framework of EMSG. However, the investigation of URs for PNSs in the context of EMSG remains unexplored. This study aims to explore four types of URs on the tidal deformability ($\Lambda$), compactness ($C$), $f$-mode frequency multiplied with canonical mass ($f_fM_{1.4}$), and gravitational binding energy per mass ($B/M$) of PNSs by varying the $\alpha$ parameter.
\\
The residuals for all the URs are calculated by the formula,
\begin{equation}
    \Delta = \frac{|y - y_{\rm fit}|}{y_{\rm fit}},
\end{equation} 
where, $y$=$f_f M_{1.4}$ and $|B|/M$.
\subsection{$f_f M_{1.4}$-$C$ relation}
Andersson and Kokkotas \cite{Kokkotas1999} first established the correlation between $C$ and $f$-mode frequency. Afterthat, in \cite{Sotani_2021}, Sotani and Kumar introduced the UR for the $f$-mode frequency multiplied by the normalized neutron star mass ($f_f M_{1.4}$). Here, we calculate the $f_f M_{1.4}$-$C$ relations for PNSs in EMSG, using the approximate formula obtained through least-squares fitting,
\begin{equation}
\label{eq:C-f_fitting}
    f_f M_{1.4} = \sum_{n=0}^{n=5} a_n (C)^n \, .
\end{equation}
We enumerated the fitting coefficients ($a_n$) and their corresponding reduced chi-squared ($\chi_r^2$) error values in Table \ref{f-c}. In the upper panels of Fig. \ref{URf}, the $f_f M_{1.4}$ is plotted as a function of $C$, for four different EOSs with the variation of $\mathrm{S}$ and $Y_l$ with a fixed value of $\alpha=+5.01$ (\textit{Left}), $\alpha=0$ (\textit{Middle}) and $\alpha=-5.01$ (\textit{Right}). The residuals ($\Delta$) are also plotted in the lower panel of each figure.

\subsection{$f_f M_{1.4}$-$\Lambda$ relation}
\begin{figure*}
    \includegraphics[width=0.55\textwidth, angle=-90]{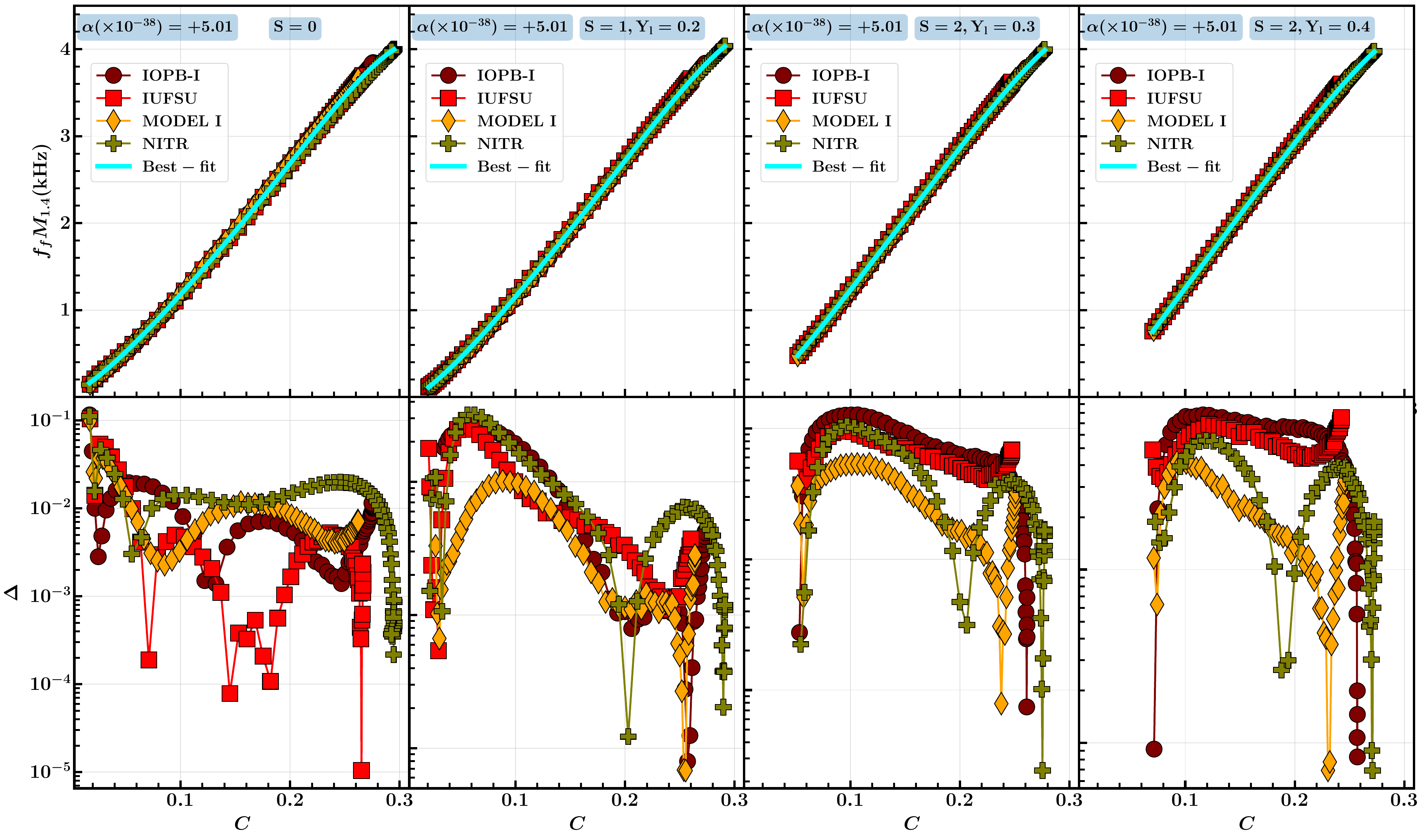}
    \includegraphics[width=0.55\textwidth, angle=-90]{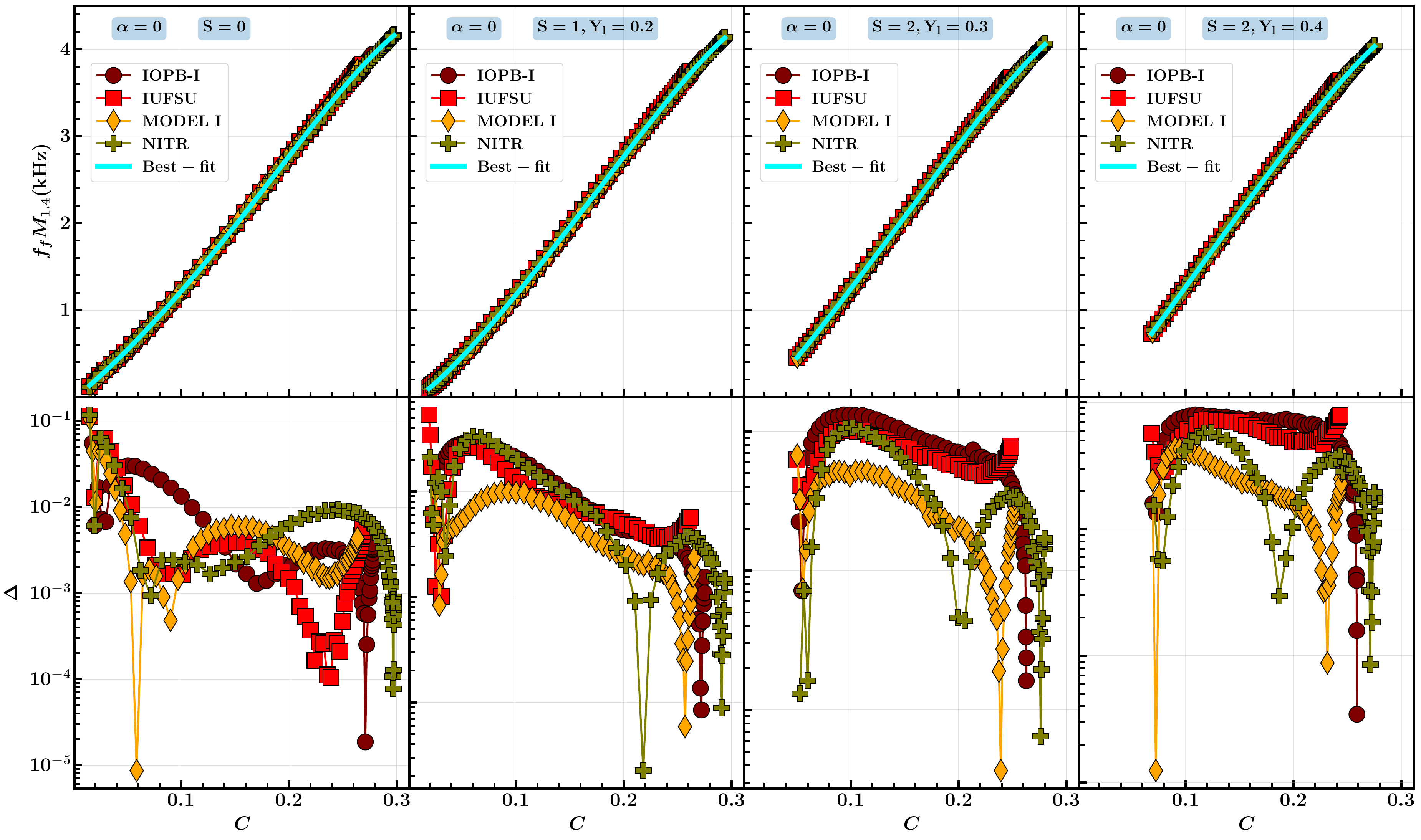}
    \includegraphics[width=0.55\textwidth, angle=-90]{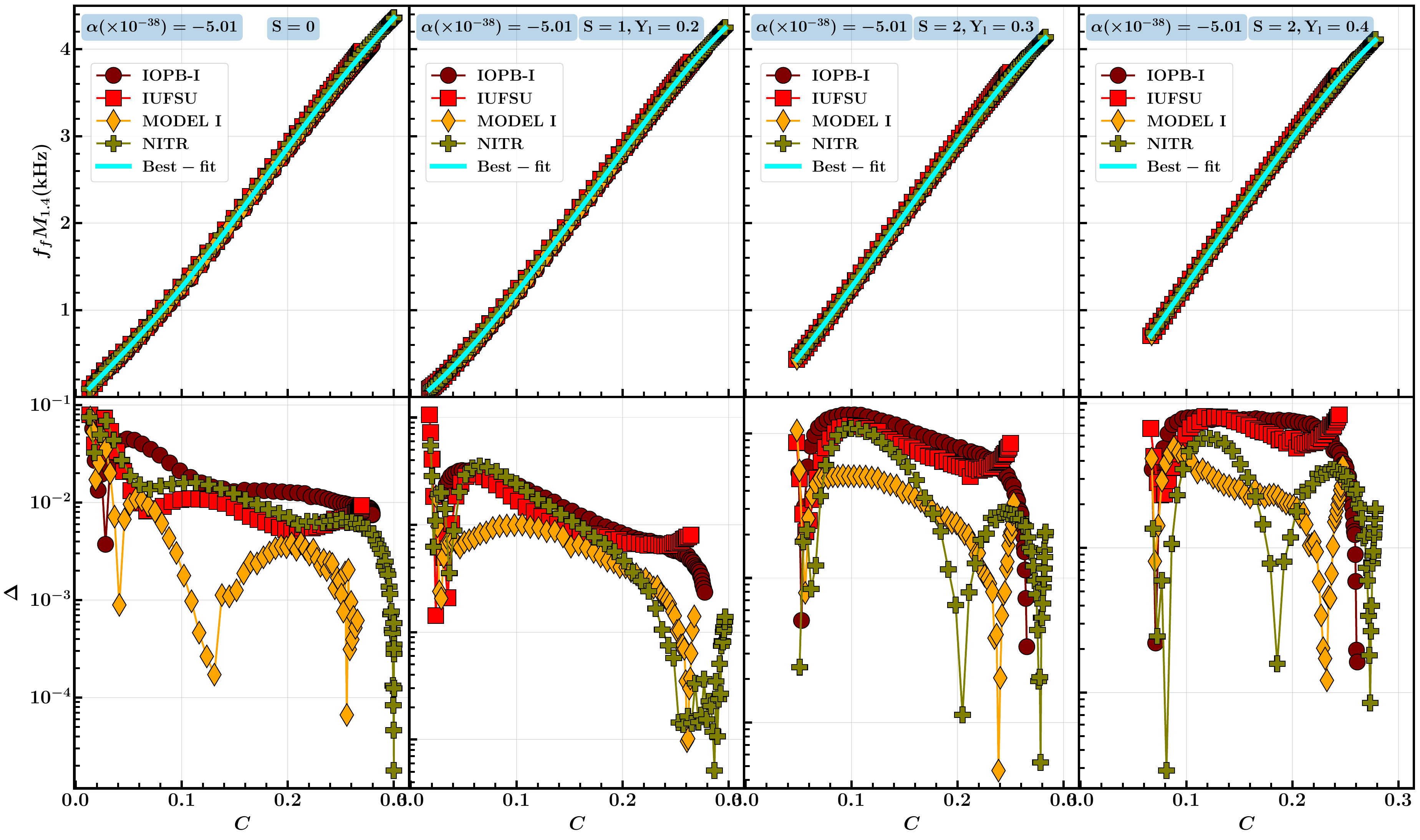}
    \includegraphics[width=0.55\textwidth, angle=-90]{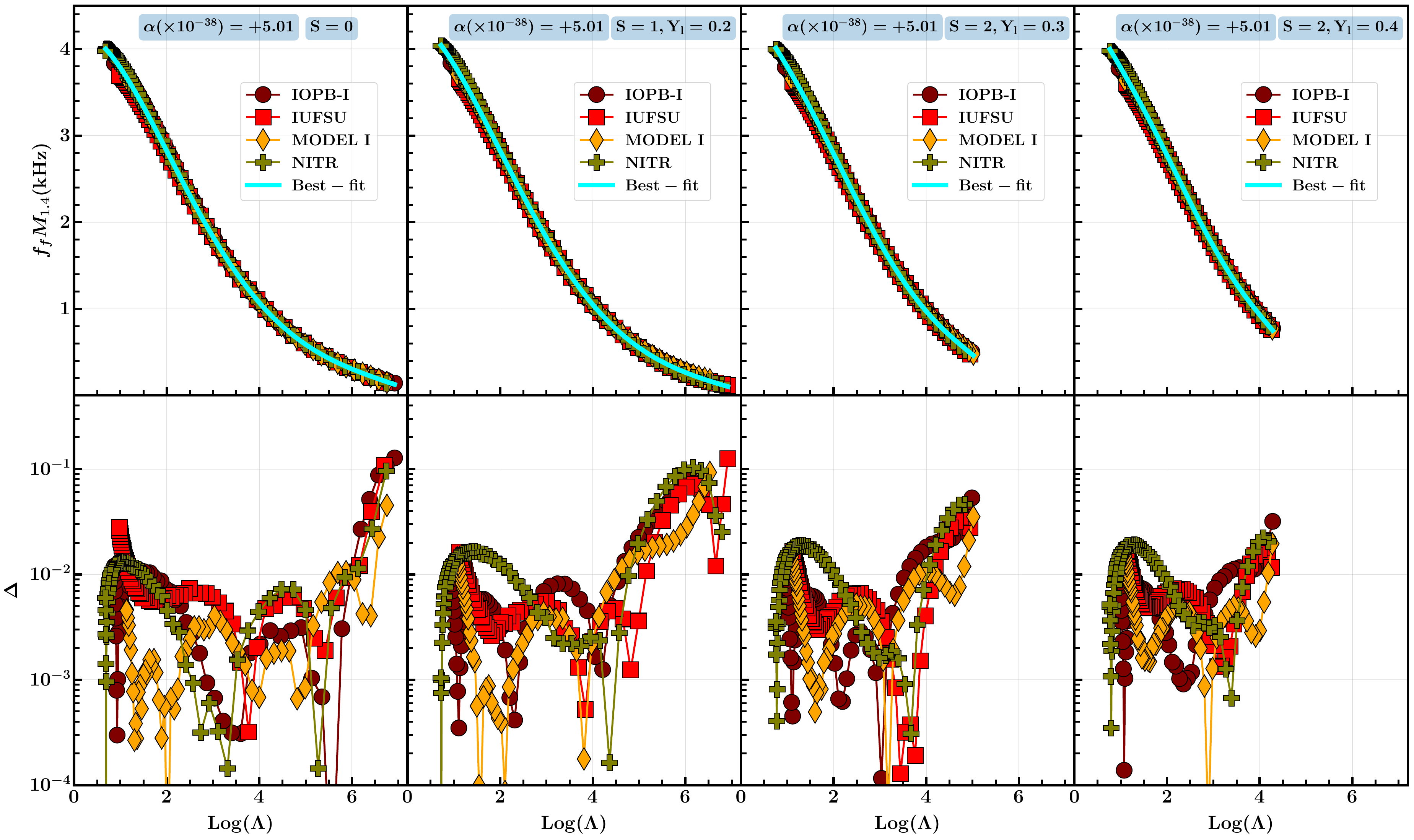}
    \includegraphics[width=0.55\textwidth, angle=-90]{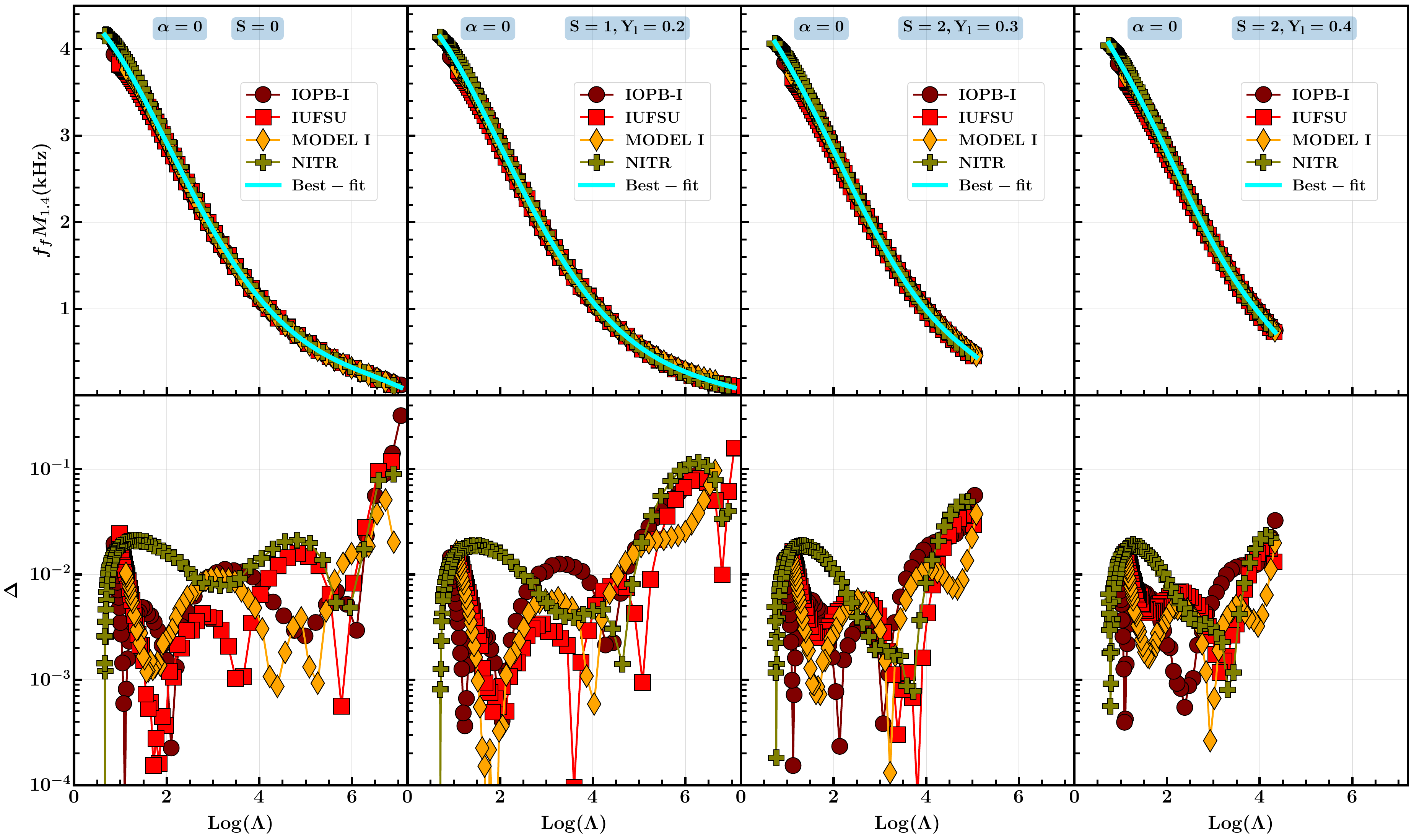}
    \includegraphics[width=0.55\textwidth, angle=-90]{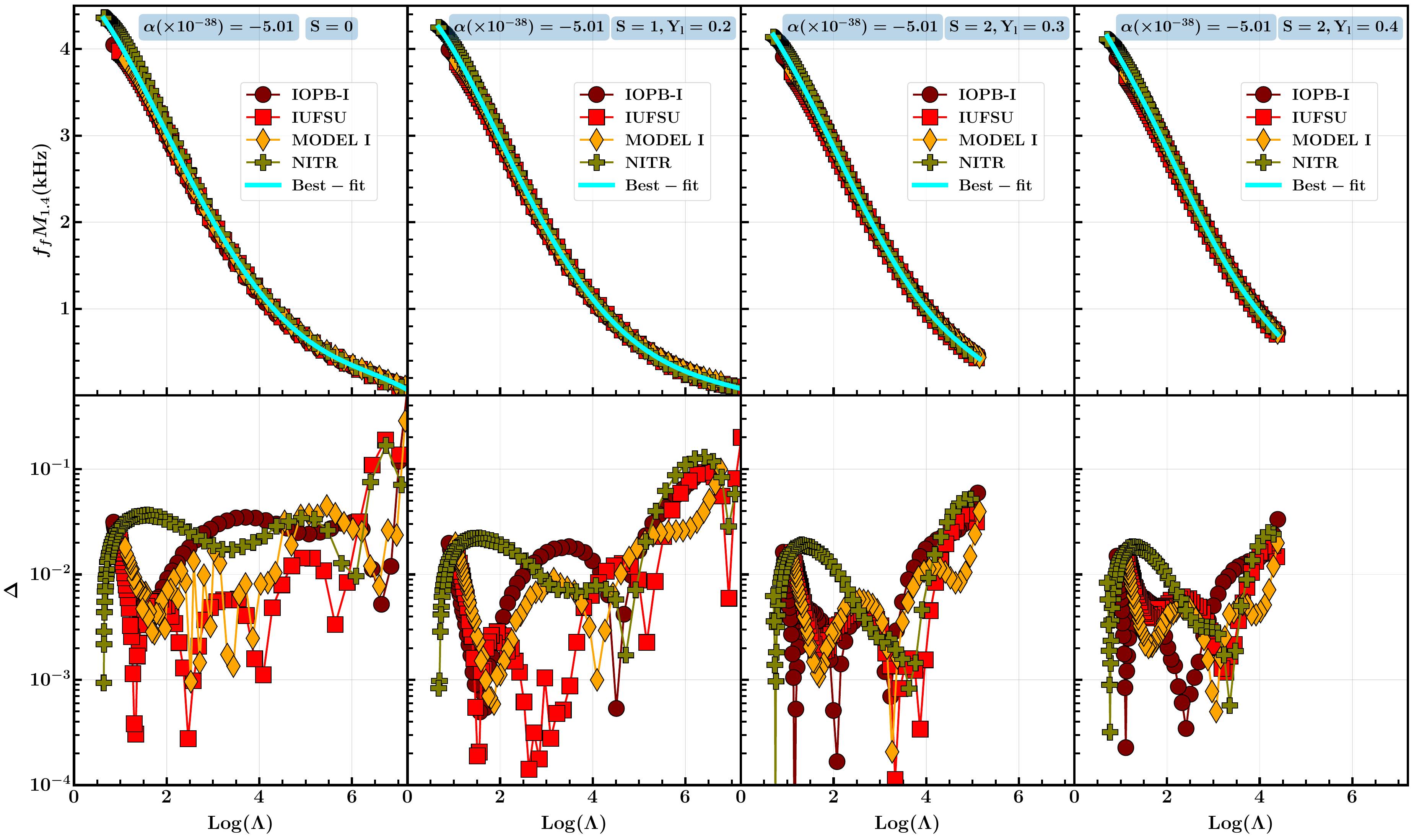}
    \caption{$f_f M_{1.4}$-$C$ (\textit{upper}) and $f_f M_{1.4}$-$\Lambda$ (\textit{lower}) relations for four different EOSs with the variation of $\mathrm{S}$ and $Y_l$ with a fixed value of $\alpha=+5.01$ (\textit{left}), $\alpha=0$ (\textit{middle}) and $\alpha=-5.01$ (\textit{right}). Each of the lower panels of the six figures contains the plot for residuals ($\Delta$).}
    \label{URf}
\end{figure*}

According to \cite{Sotani_2021}, here, we calculate the $f_f M_{1.4}$-$\Lambda$ relations for PNSs in EMSG, using the approximate formula obtained through least-squares fitting
\begin{equation}
\label{eq:C-f_fitting}
    f_f M_{1.4} = \sum_{n=0}^{n=5} b_n (\Lambda)^n \, .
\end{equation}
We enumerated the fitting coefficients ($b_n$) and their corresponding reduced chi-squared ($\chi_r^2$) error values in Table \ref{f-Love}. In the lower panels of Fig. \ref{URf}, the $f_f M_{1.4}$ is plotted as a function of $\Lambda$.
\subsection{$|B|/M$-$C$ relation}
\begin{figure*}
    \includegraphics[width=0.55\textwidth, angle=-90]{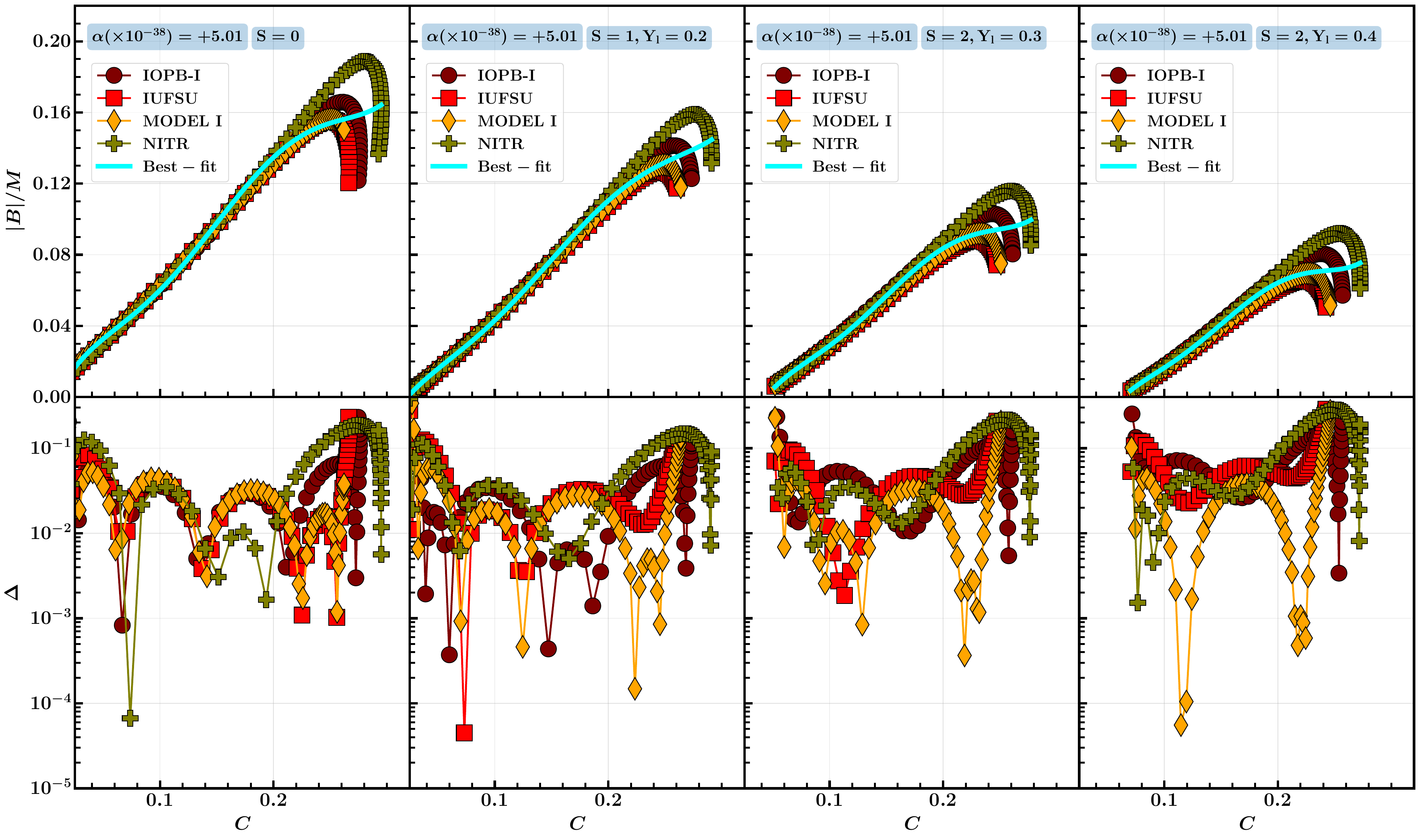}
    \includegraphics[width=0.55\textwidth, angle=-90]{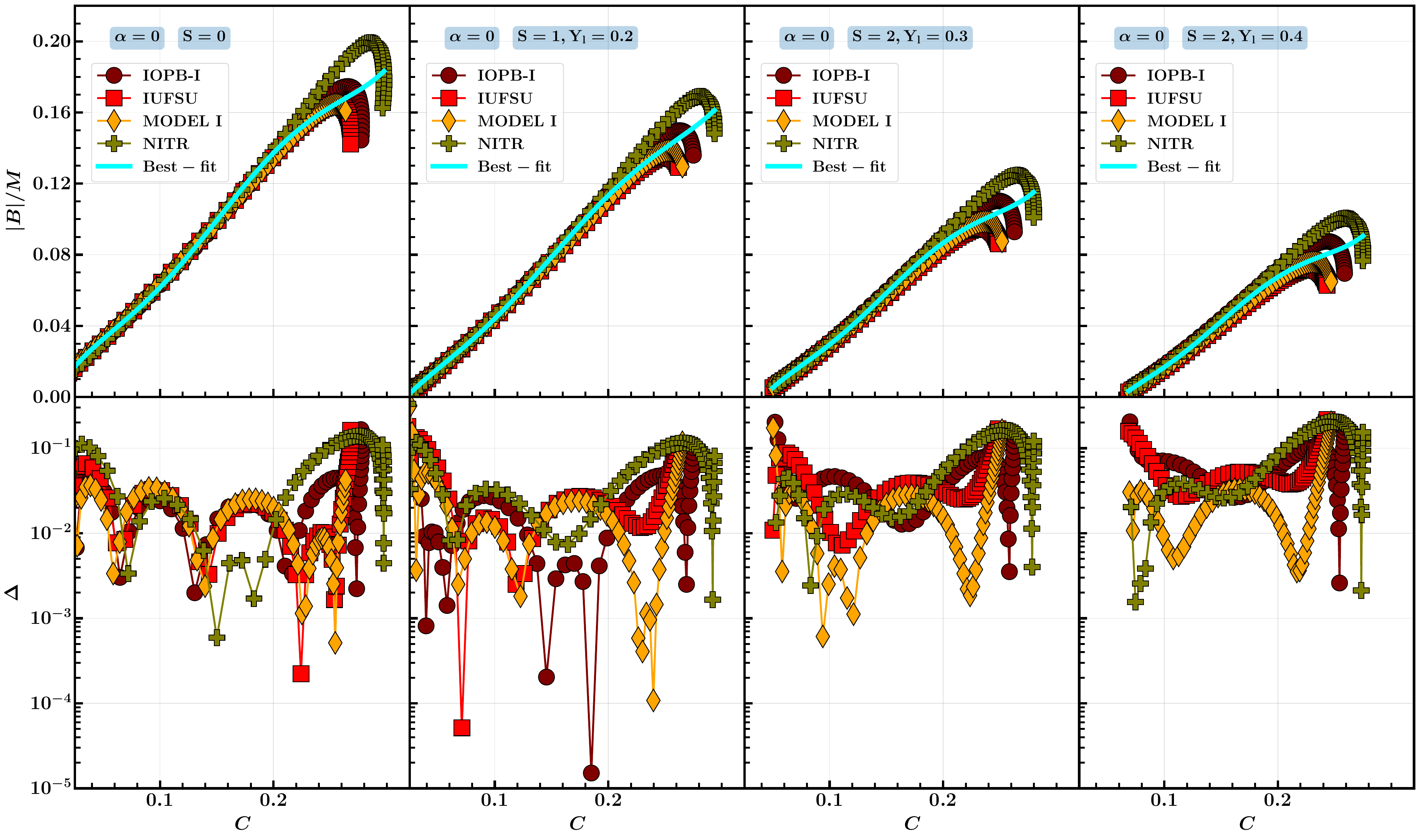}
    \includegraphics[width=0.55\textwidth, angle=-90]{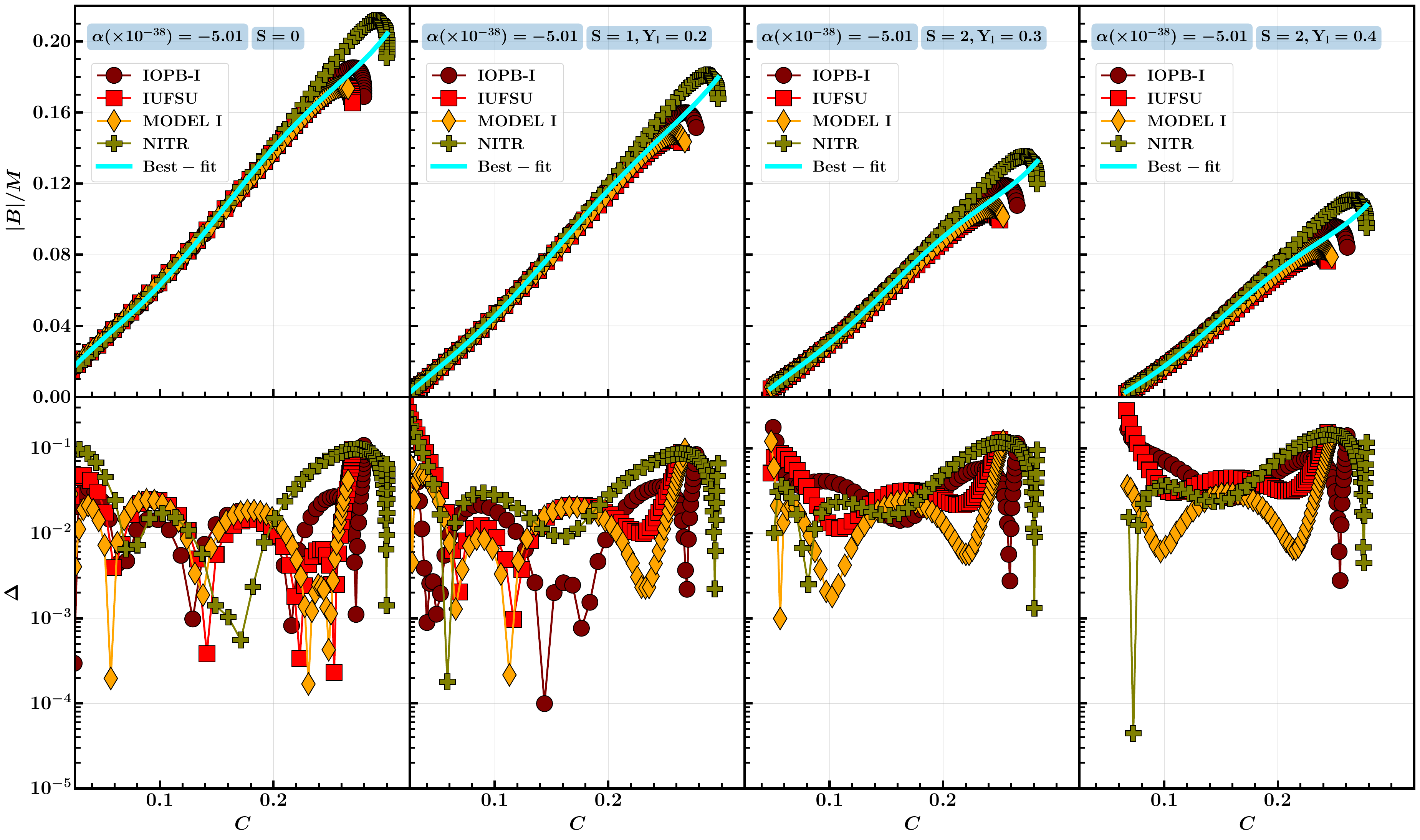}
    \includegraphics[width=0.55\textwidth, angle=-90]{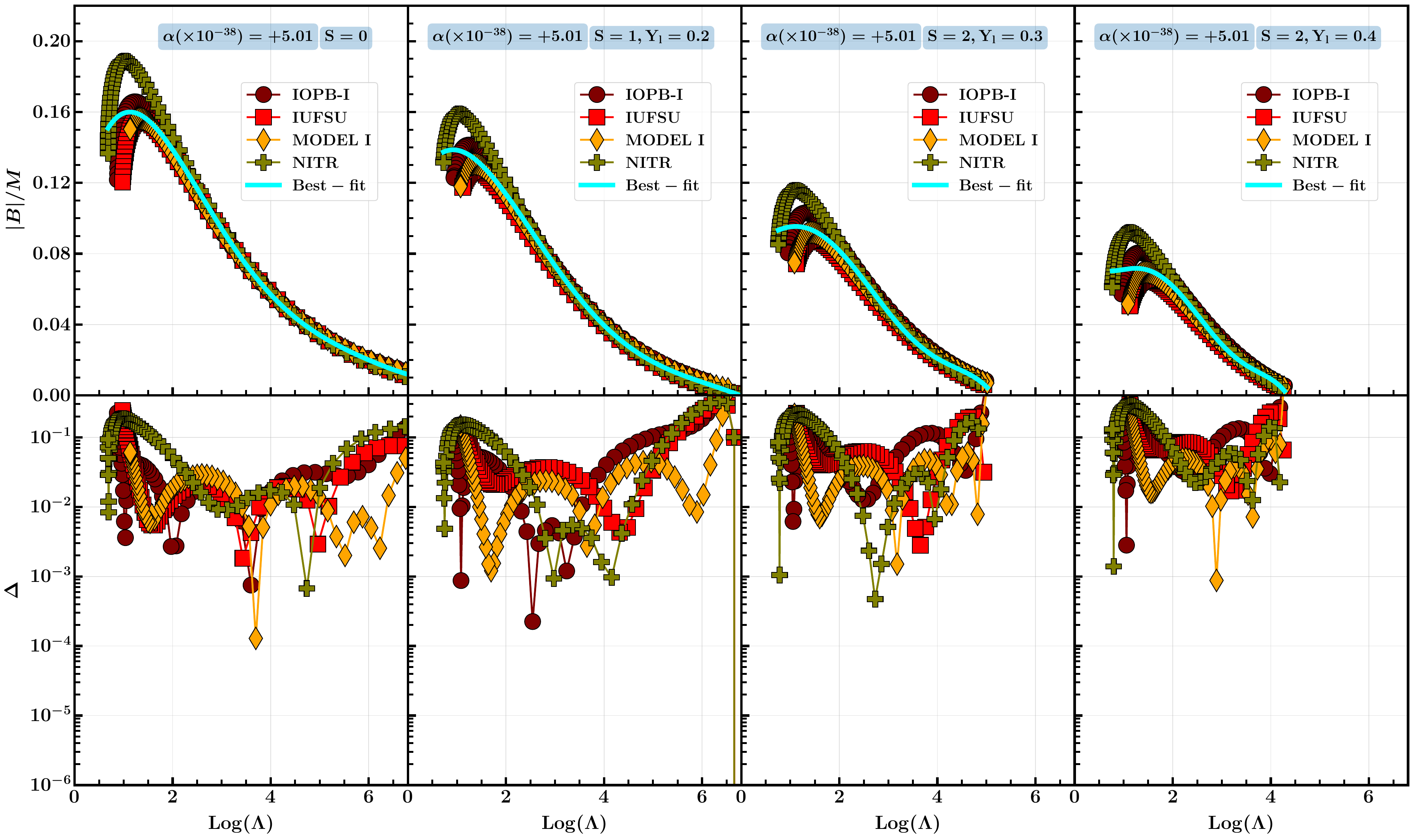}
    \includegraphics[width=0.55\textwidth, angle=-90]{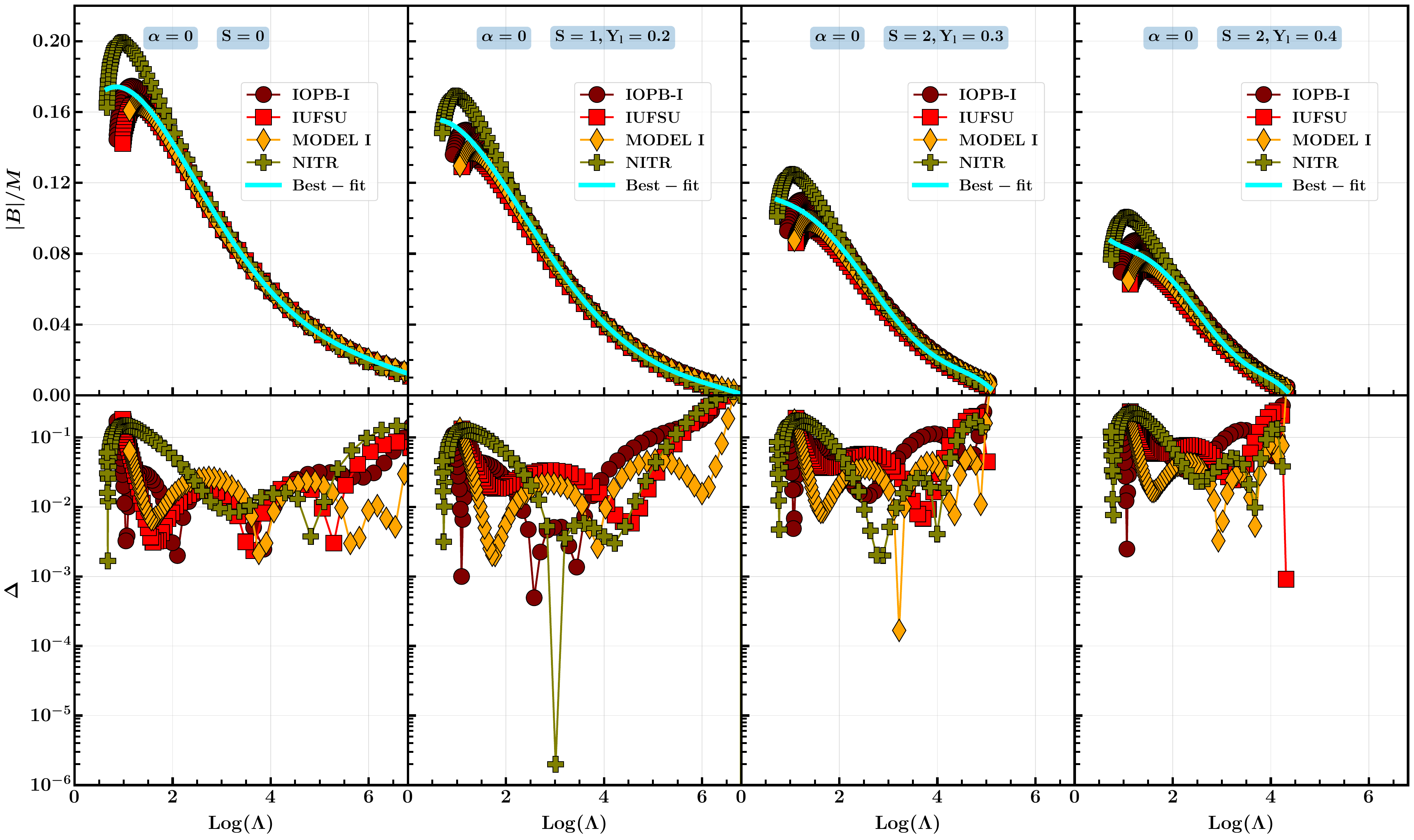}
    \includegraphics[width=0.55\textwidth, angle=-90]{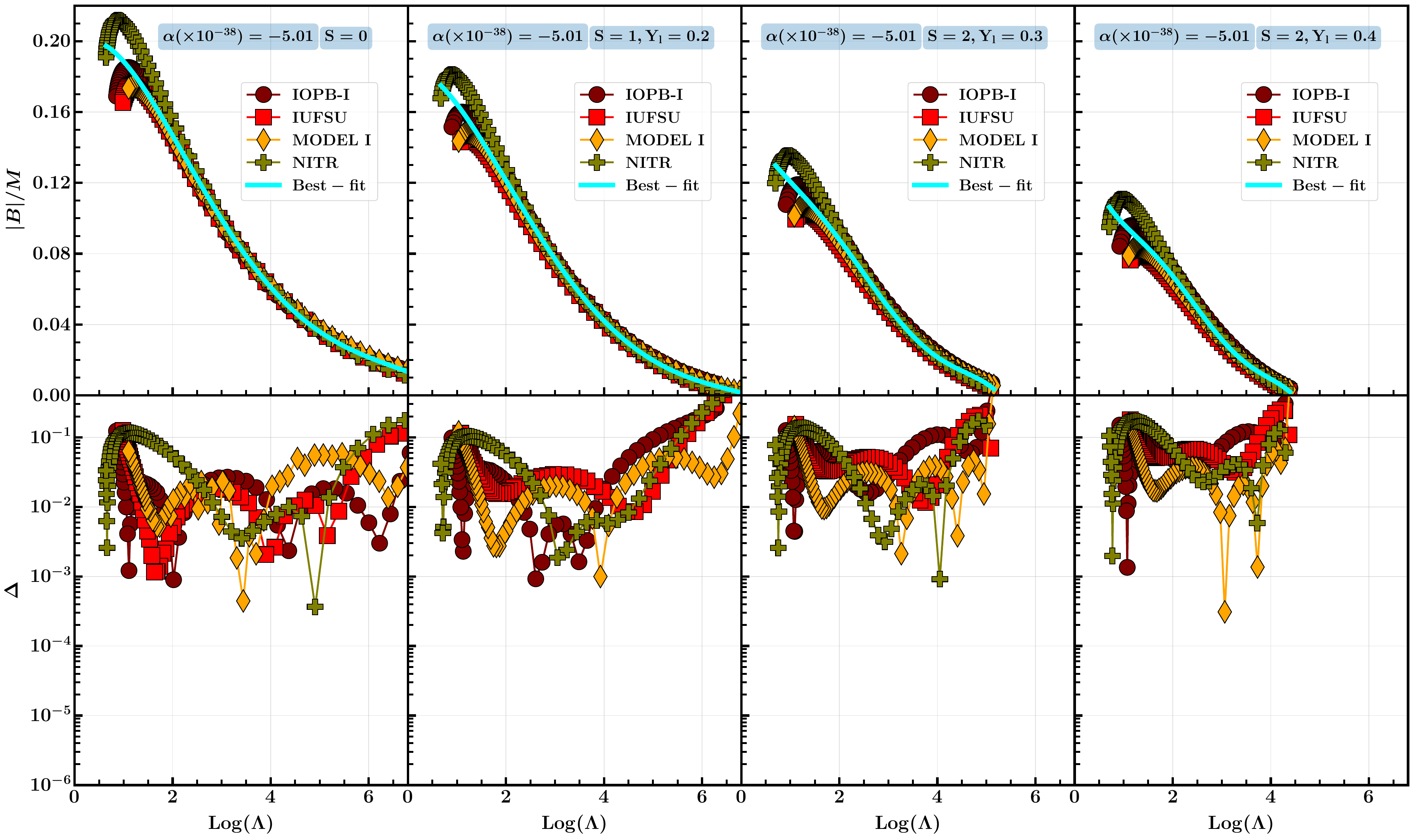}
    \caption{Same as Fig. \ref{URf}, but for $|B|/M$-$C$ and $|B|/M$-$\Lambda$ relations.}
    \label{URBE}
\end{figure*}

Followed by \cite{PhysRevD.105.063023}, we compute the $|B|/M$-$C$ relations but for PNSs in EMSG, using the approximate formula obtained through least-squares fitting
\begin{equation}
\label{eq:C-f_fitting}
    |B|/M = \sum_{n=0}^{n=5} c_n (C)^n \, .
\end{equation}
We enumerated the fitting coefficients ($c_n$) and their corresponding reduced chi-squared ($\chi_r^2$) error values in Table \ref{B-C}. In the upper panels of Fig. \ref{URBE}, the $|B|/M$ is plotted as a function of $C$.


\subsection{$|B|/M$-$\Lambda$ relation}
We now evaluate the $|B|/M$-$\Lambda$ relations for PNSs in EMSG, using the approximate formula obtained through least-squares fitting
\begin{equation}
\label{eq:C-f_fitting}
    |B|/M = \sum_{n=0}^{n=5} d_n (\Lambda)^n \, .
\end{equation}
We enumerated the fitting coefficients ($d_n$) and their corresponding reduced chi-squared ($\chi_r^2$) error values in Table \ref{B-Love}. In the lower panels of Fig. \ref{URBE}, the $|B|/M$ is plotted as a function of $\Lambda$.
\section{Correlation of universal relations}
\label{sec:CorrUR}
\begin{figure*}
    \includegraphics[width=\textwidth]{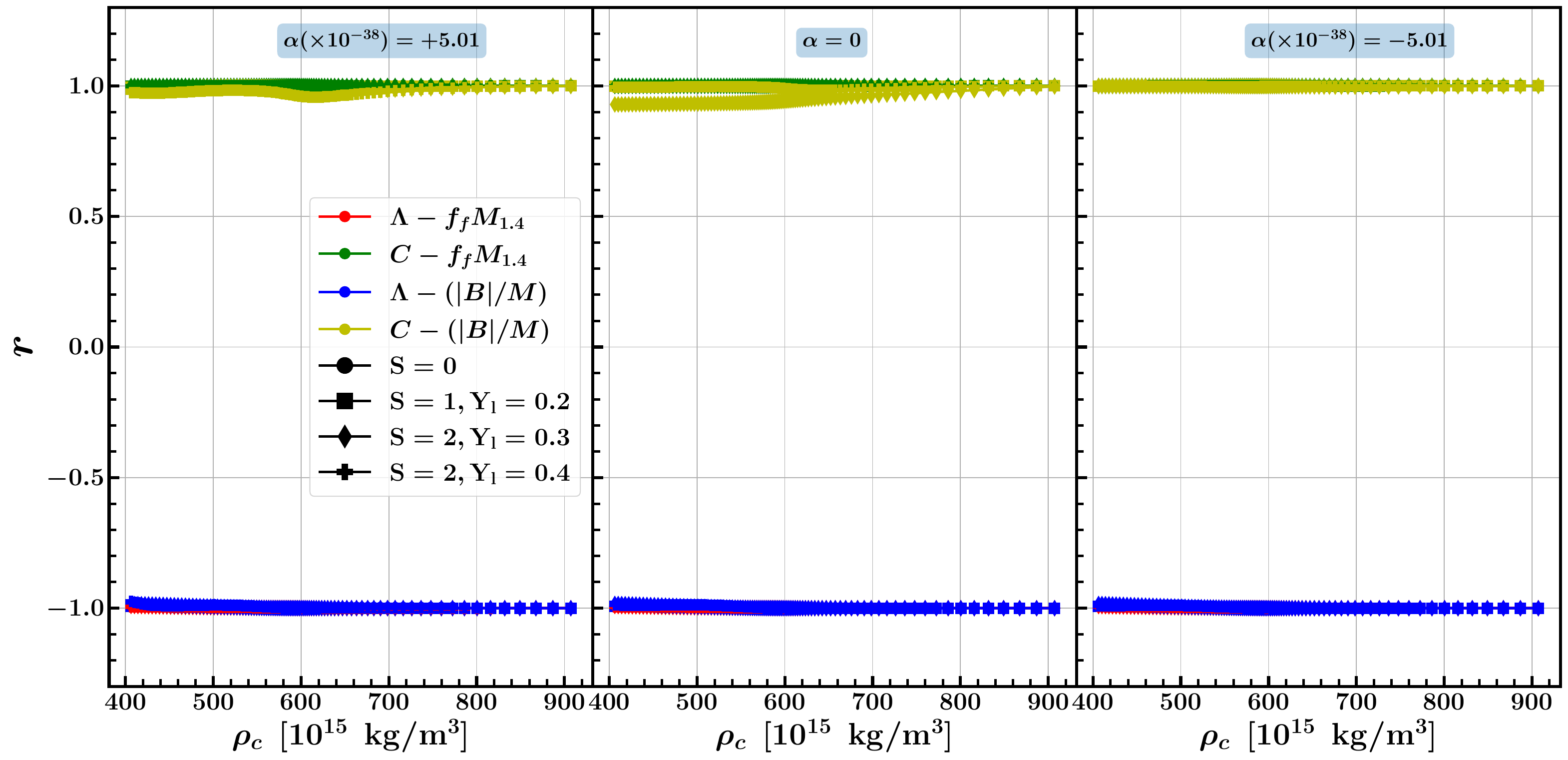}
    \caption{Linear Correlation coefficient ($r$) of the four different universal relations (represented by different colours) for different PNSs by varying the central density ($\rho_c$). Different markers defined different values of $S$ and $Y_l$. 
    Three panels are for three different values of $\alpha$. $\alpha=+5.01$ (\textit{Left}), $\alpha=0$ (\textit{Middle}), and $\alpha=-5.01$ (\textit{Right}).}
    \label{Corr}
\end{figure*}
\begin{figure*}
    \includegraphics[width=\textwidth]{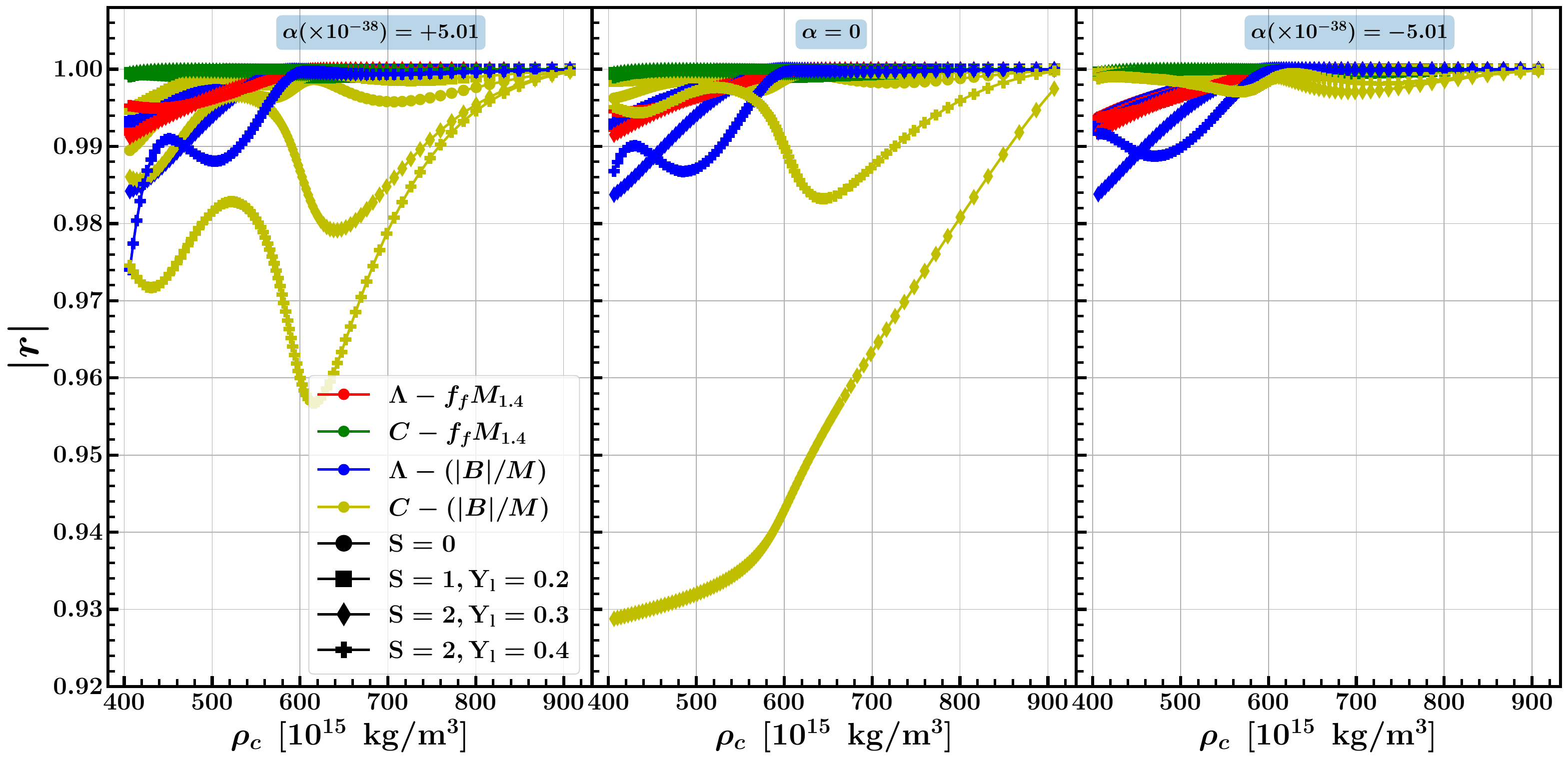}
    \caption{Same as Fig. \ref{Corr} but with the absolute values of the correlation coeffieients ($|r|$).}
    \label{Corr_abs}
\end{figure*}
Universal relations provide an effective way to study the connections among global properties of NSs \cite{YAGI20171, PhysRevD.100.123010, PhysRevD.102.023039}. These relations arise from strong correlations between different global quantities \cite{PhysRevD.100.123010, PhysRevD.102.023039, PhysRevD.101.043021}. Therefore, exploring such correlations helps to identify new universal relations. In this context, linear correlation analysis is a useful tool for quantifying the relationship between two quantities, with the correlation coefficient $r$ measuring the strength of their linear dependence, expressed as
\begin{equation}
    r(X,Y)=\frac{\mathrm{Cov}(X,Y)}{\sqrt{D(X)} \sqrt{D(Y)}},
\end{equation}
where $X$, $Y$ represent the object variables, $\mathrm{Cov}$($X$,$Y$) is the covariance, and $D(X)$, $D(Y)$ represent the variance of $X$, $Y$ respectively \cite{PhysRevD.101.043021, PhysRevC.102.052801, PhysRevD.105.063023}. The closer the absolute value of the correlation coefficient, $|r|$, is to 1, the stronger and more nearly linear the correlation between the two quantities. Conversely, as $|r|$ approaches 0, the correlation strength becomes weaker.
\\
To quantify EOS-independence, we computed the
correlation coefficients for four different URs, such as $f_fM_{1.4}$-$C$, $f_fM_{1.4}$-$\Lambda$, $|B|/M$-$C$ and $|B|/M$-$\Lambda$. In Fig. \ref{Corr}, we present the correlation coefficient values ($r$) for the four mentioned URs, evaluated for different values of $S$, $Y_l$ and $\alpha$ across various PNSs by varying the central density ($\rho_c$). We find that the correlation of $f_fM_{1.4}$ and $|B|/M$ with $C$ is positive, and negative with $\Lambda$. That is $f_fM_{1.4}$ and $|B|/M$ increases as $C$ increases and $\Lambda$ decreases. Irrespective of the negative and positive values, we find that the correlation is almost 1. 
\\
To quantify the strength of the correlation, we consider the absolute value of the correlation coefficient $|r|$, as shown in Fig. \ref{Corr_abs}. It is clearly visible that the $|r|$ ranges between 0.9 and 1. However, small variations in $|r|$ are observed when changing $S$, $Y_l$, and $\alpha$. Despite these differences, the correlations remain strong in all cases. This indicates that although macroscopic properties such as mass, radius, frequency, tidal deformability may vary with $S$, $Y_l$, and $\alpha$, the correlation of the URs of these macroscopic properties remains significantly unchanged by these variations. As a result, correlation analyses provide a powerful framework for constraining PNSs' structure, interpreting astrophysical observations (e.g., gravitational waves), and testing gravity theories, even when detailed information about the internal composition of PNSs is not directly accessible.
\section{Summary and Conclusions}
\label{sec:summary}
In this work, we presented a comprehensive analysis of proto-neutron star properties within the framework of Energy-Momentum Squared Gravity (EMSG) using four different realistic finite-temperature RMF EOS models (NITR, IOPB-I, MODEL I, and IUFSU). We have verified the causality conditions for all the EOSs by varying the EMSG coupling parameter $\alpha$. First, we generated the finite temperature EOSs, each of which contains four different combinations of $S$ and $Y_l$ ($S=0$ for NSs) and took the temperature-dependent crust part from COMPOSE. We solved the modified TOV equations (modified by EMSG) for the four mentioned finite temperature EOSs. We obtained the mass-radius relation for different combinations of $S$ and $Y_l$, and the bands of $M-R$ for the range of $\alpha$. To visualise how the effect of $\alpha$ varied for different combinations of $S$ and $Y_l$, we have calculated the bandwidth of each $M-R$ curve by fixing $M$, i.e. for different values of $M$, we took the difference of $R$ ($\Delta R$) for the highest positive and lowest negative values of $\alpha$. We found that the PNSs with higher values of $S$ and $Y_l$ affected more, i.e. the $\Delta R$ is more.
\\
We then computed the non-radial $f$-mode oscillation to understand how $S$, $Y_l$ and $\alpha$ affect it. Similarly to the $M-R$ relation, here also we have calculated the bandwidth for frequency ($\Delta f_f$) by fixing the masses of PNSs. We found that the $\Delta f_f$ is less for $S=2, Y_l=0.4$ compared to all lower combination values of $S$ and $Y_l$. It is obvious that the radius bandwidth is more for $S=2, Y_l=0.4$, and we know that for fixed mass, compactness ($C=\frac{M}{R}$) will be less, which gives weaker gravity, meaning lower restoring force, and the frequency will decrease. We got that for $S=2, Y_l=0.4$, frequency band width is less.
\\
After that, we calculated the effect of $S$, $Y_l$ and $\alpha$ on Gravitational binding energy. We found that the effect of $\alpha$ on binding energy is more, near the higher values of mass of the PNSs. We found that $\Delta (B/M)$, the width of binding energy per mass of the PNSs, is less for $S=2, Y_l=0.4$, compared to the other lower combinations of $S$ and $Y_l$. It is also obvious asper our previous discussion. Since with increasing radius, gravity gets weaker, i.e., the PNSs become less tightly held together by gravity, so the binding energy decreases.
\\
After calculating these macroscopic properties of the PNSs, for the variation of $S$, $Y_l$, and $\alpha$ values, we found that there are visible effects of them on the PNSs. However, to make an EOS independent study, we then focus on universal relations which were not explored in any previous study for all the variations of $S$, $Y_L$, and $\alpha$ parameters. We focused on four different URs, such as $f_f M_{1.4}$-$C$, $f_f M_{1.4}$-$\Lambda$, $|B|/M$-$C$, and $|B|/M$-$\Lambda$. We have calculated the fitting coefficients and their corresponding reduced chi-squared ($\chi_r^2$) error values for the mentioned four URs. We found that the variation in $S$, $Y_l$, and $\alpha$ values does not effectively change the power of reduced chi-squared ($\chi_r^2$) error, indicating a stronger EOS-insensitive relation and vice versa. 
\\
Now, to quantify the strong UR, we calculated the correlation coefficients for the four mentioned URs. We found that, $f_f M_{1.4}$ and $|B|/M$ are positively correlated with $C$, and negatively correlated with $\Lambda$. To know how strongly they are correlated, we calculated the absolute value of the correlation coefficients. However, we found that all the correlation coefficients are within the range of 0.92 to 1, which means strong correlation, for the variation of $S$, $Y_l$ and $\alpha$. So, we can conclude by saying that, however, the macroscopic properties such as mass, radius, $f$-mode, tidal deformability, compactness, etc., are affected by the variation of $S$, $Y_l$ and $\alpha$, but the universal relations and the correlations of the universal relations remain strong and largely unaffected.
\section{Acknowledgments}
I would like to thank my supervisor Dr. Bharat Kumar, for his helpful suggestions. I also want to thank Dr. Suprovo Ghosh for the fruitful discussions during the COAA conference at IIT Jodhpur. 
\bibliographystyle{APP} 
\bibliography{APP}

\end{document}